\documentclass[aps,floatfix,showpacs,nofootinbib,,amsmath,amssymb]
{revtex4}
\usepackage[dvips]{graphicx}

\begin{document}

\newcommand{\Frac}[2]{\frac
{\begin{array}{@{}c@{}}\strut#1\strut\end{array}}
{\begin{array}{@{}c@{}}\strut#2\strut\end{array}}}


\title{On cosmological observables in a swiss-cheese universe}

\author{Valerio Marra} \email{valerio.marra@pd.infn.it}

\affiliation{Dipartimento di Fisica ``G.\ Galilei'' Universit\`{a} di Padova,
INFN Sezione di Padova, via Marzolo 8, Padova I-35131, Italy}
\affiliation{Department of
Astronomy and Astrophysics, the University of Chicago, 
Chicago, IL \ \ 60637-1433 }

\author{Edward W.\ Kolb} \email{rocky.kolb@uchicago.edu} 
\affiliation{ Department of Astronomy and Astrophysics, Enrico Fermi
Institute, and  Kavli Institute for Cosmological Physics, the University of
Chicago, Chicago, IL \ \ 60637-1433 }

\author{Sabino Matarrese} \email{sabino.matarrese@pd.infn.it}
\affiliation{Dipartimento di Fisica ``G.\ Galilei'' Universit\`{a} di Padova,
INFN Sezione di Padova, via Marzolo 8, Padova I-35131, Italy}

\author{Antonio Riotto} \email{antonio.riotto@pd.infn.it}
\affiliation{D\'epartement de Physique Th\'eorique, Universit\'e de G\`eneve, 
24 Quai Ansermet, G\`eneve, Switzerland, and
INFN Sezione di Padova, via Marzolo 8, Padova I-35131, Italy}

\begin{abstract}

Photon geodesics are calculated in a swiss-cheese model, where the cheese is
made of the usual Friedmann-Robertson-Walker solution and the holes are
constructed from a Lema\^itre-Tolman-Bondi solution of Einstein's equations.  
The observables on which we focus are the changes in the redshift, in the
angular-diameter--distance relation, in the luminosity-distance--redshift
relation, and in the corresponding distance modulus. We find that redshift
effects are suppressed when the hole is small because of a compensation effect
acting on the scale of half a hole resulting from the special case of spherical
symmetry.  However, we find interesting effects in the calculation of the
angular distance: strong evolution of the inhomogeneities (as in the approach
to caustic formation) causes the photon path to deviate from that of the FRW
case.  Therefore, the inhomogeneities are able to partly mimic the effects of a
dark-energy component. Our results also suggest that the nonlinear effects of
caustic formation in cold dark matter models may lead to interesting effects on
photon trajectories. 

\end{abstract}


\maketitle


\section{Introduction}

In this paper we explore a toy cosmological model in order to attempt to
understand the role of large-scale non-linear cosmic inhomogeneities in the
interpretation of observable data. The model is based on a swiss-cheese model,
where the cheese consists of the usual Friedmann-Robertson-Walker (FRW)
solution and the holes are constructed out of a Lema\^{i}tre-Tolman-Bondi (LTB)
solution.  The advantage of this model is that it is a solvable model with strong
nonlinearities, in particular the formation of caustics as expected in the cold
dark matter (CDM) models.

Most, if not all, observations are consistent with the cosmic concordance model
according to which, today, one-fourth of the mass-energy of the universe is
clustered and dominated by cold dark matter.  The remaining three-quarters is
uniform and dominated by a fluid with a  negative pressure (dark energy, or
$\Lambda$). 

While the standard $\Lambda$CDM model seems capable of accounting for the
observations, it does have the feature that approximately 95\% of the
mass-energy of the present universe is unknown. We are either presented with
the opportunity of discovering the nature of dark matter and dark energy, or
nature might be different than described by the $\Lambda$CDM model. Regardless,
until such time as dark matter and dark energy are completely understood, it is
useful to look for alternative cosmological models that fit the data.  

One  non-standard possibility is that there are large effects on
the {\it observed} expansion rate due to the back reaction of inhomogeneities
in the universe.  The basic idea is that all evidence for dark energy comes
from observational determination of the expansion history of the universe. 
Anything that affects the observed expansion history of the universe alters the
determination of the parameters of dark energy; in the extreme it may remove
the need for dark energy.

This paper focuses on the effects of large-scale nonlinear inhomogeneities on
observables such as the luminosity-distance--redshift relation. The ultimate
goal is to find a realistic dust model that can explain observations (like the
luminosity-distance--redshift relation) without the need of dark energy. The
ultimate desire will be to have an exactly solvable realistic inhomogeneous
model.  Our model is but a first small step in this pragmatic and necessary
direction.

If this first step is successful, we would show that inhomogeneities must be
factored into the final solution.  Even if we live in a $\Lambda$CDM universe,
inhomogeneities ``renormalize'' the theory adding an effective extra  source to
the dark energy. We have to be very careful in what we mean. The
inhomogeneities renormalize the dust Einstein--de Sitter universe only from the
observational point of view, that is luminosity and redshift of photons.
Average dynamics is beyond this and we will not be concerned with this issue 
in this paper: if we find an effective cosmological constant, this will 
not mean that the universe is accelerating, but only that its 
luminosity distance-redshift relation will fit the observational data. 

Here we are not primarily interested in the backreaction effect  that comes
from the  averaging procedure in General Relativity (see e.g.
\cite{buchert_new}).  Since we have an exact solution, we can directly
calculate observables. Indeed, in this paper we are mainly interested in the
effect of inhomogeneities on the dynamics of photons. 

We can reformulate our present approach as follows:  inhomogeneities
renormalize the geodesics of photons.  In the extreme case in which such a
renormalization leads to a negative effective deceleration parameter in the
luminosity distance-redshift relation, it might make us think that a
dark-energy component exists. 

The paper is organized as follows: In Sect.\ \ref{model} we will specify the
parameters of our swiss-cheese model.  In Sect.\ \ref{dynamics} we study its
dynamics.  Then in Sect.\ \ref{photons} we will discuss the geodesic equations
for light propagation. We will apply them to see what an observer in the cheese
(Sect.\ \ref{cheese}) or in the hole (Sect.\ \ref{hole}) would observe. The
observables on which we will focus are the change in redshift $\Delta z$,
angular-diameter--distance $\Delta d_{A}(z)$, luminosity-distance--redshift
relation $\Delta d_{L}(z)$, and the distance modulus $\Delta m(z)$.

Conclusions are given in Sect.\ \ref{conclusions}. In two appendices we discuss
the role of arbitrary functions  in LTB models (Appendix A) and some technical
issues in the  solution of photon geodesics in our swiss-cheese model (Appendix
B).

\section{The model} \label{model}

We study a swiss-cheese model where the cheese consists of the usual
Friedmann--Robertson--Walker solution and the spherically symmetric holes
are constructed from a Lema\^{i}tre-Tolman-Bondi solution.  The
particular FRW solution we will choose is a matter-dominated, spatially-flat
solution, \textit{i.e.,} the Einstein--de Sitter (EdS) model.

In this section we will describe the FRW and LTB model parameters we have
chosen.  But first, in Table \ref{units} we list the units we will use for mass
density, time, the radial coordinate, the expansion rate, and two quantities,
$Y(r,t)$ and $W(r)$, that will appear in the metric.

The time $t$ appearing in Table \ref{units} is not the usual time in FRW
models.  Rather, $t=T-T_0$, where $T$ \textit{is} the usual cosmological time
and $T_0=2H_0^{-1}/3$ is the present age of the universe.  Thus, $t=0$ is the
present time and $t=t_{BB}=-T_0$ is the time of the big bang.  Finally, the
initial time of the LTB evolution is defined as $\bar{t}$. 

Both the FRW and the LTB metrics can be written in the form
\begin{equation}
ds^2 = -dt^{2}+\frac{Y'^2(r,t)}{W^2(r)}dr^2+Y^2(r,t) \, d\Omega^2 ,
\end{equation}
where here and throughout, the ``prime'' superscript denotes $d/dr$ and the 
``dot'' superscript will denote $d/dt$.  It is clear that the Robertson--Walker
metric is recovered with the substitution $Y(r,t)=a(t)r$ and $W^2(r)=1-kr^2$.

The above metric is expressed in the synchronous and comoving gauge. 

\begin{table}
\caption{\label{units} Units for various quantities.  We use 
geometrical units, $c=G=1$. Here, the present critical density is
$\rho_{C0}=3H^{2}_{0,\, Obs}/8 \pi$, with $H_{0,\, Obs}=70 \textrm{ km
s}^{-1}\textrm{ Mpc}^{-1}$. }
\begin{ruledtabular}
\begin{tabular}{lccr}
Quantity          & Notation    & Unit            & Value            \\ 
\hline
mass density & $\rho(r,t)$, $\bar{\rho}(r,t)$ & $\rho_{C 0}$ 
& $9.2\times10^{-30}\textrm{ g cm}^{-3}$             \\
time              & $t$, $T$, $\bar{t}$, $t_{BB}$, $T_0$ 
& $(6 \pi \rho_{C 0})^{-1/2}$  & $9.3\textrm{ Gyr}$ \\
comoving radial coordinate & $r$         & $(6 \pi \rho_{C 0})^{-1/2}$  
& $2857 \textrm{ Mpc}$ \\
metric quantity   & $Y(r,t)$    & $(6 \pi \rho_{C 0})^{-1/2}$  
& $2857 \textrm{ Mpc}$ \\
expansion rate    & $H(r,t)$    & $(6 \pi \rho_{C 0})^{1/2} $ 
& $\frac{3}{2}H_{0,\, Obs}$ \\
spatial curvature term    & $W(r)$      & $1$     &        ---             \\
\end{tabular}
\end{ruledtabular}
\end{table}

\subsection{The cheese}

We choose for the cheese model a spatially-flat, matter-dominated universe (the
EdS model).  So in the cheese there is no $r$ dependence to $\rho$ or $H$.
Furthermore, $Y(r,t)$ factors into a function of $t$ multiplying $r$ ($Y(r,t) =
a(t)r$), and in the EdS model $W(r)=1$.  In this model $\Omega_{M}=1$, so in
the cheese, the value of $\rho$ today, denoted as $\rho_0$, is unity in the
units of Table \ref{units}.   In order to connect with the LTB solution, we can
express the line element in the form
\begin{equation}
ds^2=-dt^2+Y'^2(r,t) dr^2 + Y^2(r,t) \, d\Omega^2 .
\end{equation}

In the cheese, the Friedman equation and its solution are (recall $t = 0$
corresponds to the present time):
\begin{eqnarray}
H^{2}(t) & = & \frac{4}{9} \; \rho(t)=\frac{4}{9}(t+1)^{-2} \\
Y(r,t)& = & r \, a(t)=r \, \frac{(t+1)^{2/3}}{(\bar{t}+1)^{2/3}},
\end{eqnarray}
where the scale factor is normalized so that at the beginning of the LTB
evolution it is $a(\bar{t})=1$.

For the EdS model, $T_0=1$.  We also note that the comoving distance traveled
by a photon since the big bang is $r_{BB}=3/ a_{0}$.

\subsection{The holes}

The holes are chosen to have an LTB metric \cite{lemaitre, tolman, bondi}.  The
model is based on the assumptions that the system is spherically symmetric with
purely radial motion and the motion is geodesic without shell crossing
(otherwise we could not neglect the pressure).

It is useful to define an ``Euclidean'' mass $M(r)$ and an ``average'' mass
density $\bar\rho(r,t)$, defined as
\begin{equation}
M(r) = 4\pi \int_0^r \rho(r,t) \: Y^2 Y' \: dr
= \frac{4 \pi}{3} Y^{3}(r,t) \: \bar{\rho}(r,t) .
\end{equation}
In spherically symmetric models, in general there are two expansion rates: an
angular expansion rate, $H_\perp\equiv \dot{Y}(r,t)/Y(r,t)$, and a radial
expansion rate, $H_r\equiv \dot{Y}'(r,t)/Y'(r,t)$.  (Of course in the FRW model
$H_r=H_\perp$.)
The angular expansion rate is given by
\begin{equation}
H^2_\perp(r,t) = \frac{4}{9} \; \bar{\rho}(r,t) +\frac{W^2(r)-1}{Y^{2}(r,t)}  .
\label{motion}
\end{equation}
Unless specified otherwise, we will identify $H_\perp=H$.

To specify the model we have to specify initial conditions, \textit{i.e.,} the
position $Y(r,\bar{t})$, the velocity $\dot{Y}(r,\bar{t})$ and the density
$\rho(\bar{t})$ of each shell $r$ at time $\bar{t}$. In the absence of 
shell crossing it is possible 
to give the initial conditions at different times for
different shells $r$: let us call this time $\bar{t}(r)$. The initial
conditions fix the arbitrary curvature function $W(r)$:
\begin{equation}
\label{cucu}
W^2(r)-1 \equiv 2 E(r)= \left. \left(\dot{Y}^2-  
\frac{1}{3 \pi}\frac{M}{Y}\right)\right|_{r,\bar{t}} \ ,
\end{equation}
where we can choose $Y(r,\bar{t})=r$ so that $M(r) = 4 \pi
\int_{0}^{r}\rho(\bar{r},\bar{t}) \: \bar{r}^{2} \: d\bar{r}$.

In a general LTB model there are therefore three arbitrary functions:
$\rho(r,\bar{t})$, $W(r)$ and $\bar{t}(r)$. Their values for the 
particular LTB model we study are specified in the following subsection.

In Appendix \ref{arbifunc} we provide a discussion about the number of 
independent arbitrary functions in a LTB model.

\subsubsection{Our LTB model} \label{ourmodel}

First of all, for simplicity we choose $\bar{t}(r)=\bar{t}$; \textit{i.e.,} we
specify the initial conditions for each shell at the same moment of time.

\begin{figure}
\begin{center}
\includegraphics[width=11cm]{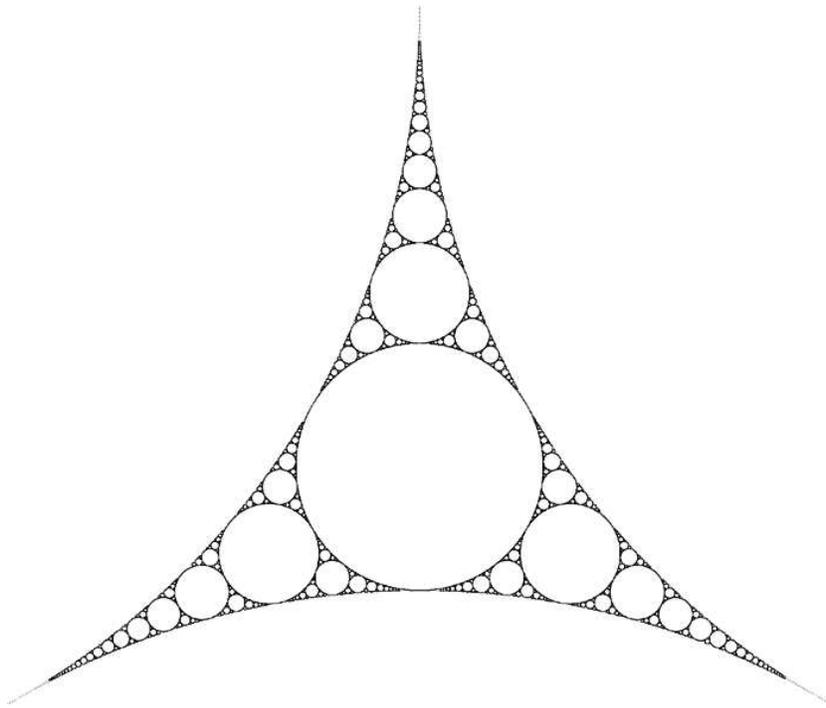}
\caption{The Apollonian Gasket.}
\label{apo}
\end{center}
\end{figure}

We now choose $\rho(r,\bar{t})$ and $W(r)$ in order to match the flat FRW model
at the boundary of the hole: \textit{i.e.,} at the boundary of the hole
$\bar{\rho}$ has to match the FRW density and $W(r)$ has to go to unity. 
A physical picture is that, given a FRW sphere, all the matter in the inner
region is pushed to the border of the sphere while the quantity of matter
inside the sphere does not change.
With the density chosen in this way, an observer outside the hole will not feel
the presence of the hole as far as \textit{local} physics is concerned (this
does not apply to global quantities, such the luminosity-distance--redshift
relation for example). So the
cheese is evolving as an FRW universe while the holes evolve differently.
In this way we can imagine putting in the cheese as many
holes as we want, even with different sizes and density profiles, and still have
an exact solution of the Einstein equations (as long as there is no
superposition among the holes and the correct matching is achieved).
The limiting picture of this procedure is
the Apollonian Gasket of Fig.\ \ref{apo}, where all the possible holes are
placed, and therefore the model has the strange property that it is FRW
nowhere, but it behaves as an FRW model on the average. This idea was first
proposed by Einstein and  Straus~\cite{einstein}.

\begin{figure}
\begin{center}
\includegraphics[width=14 cm]{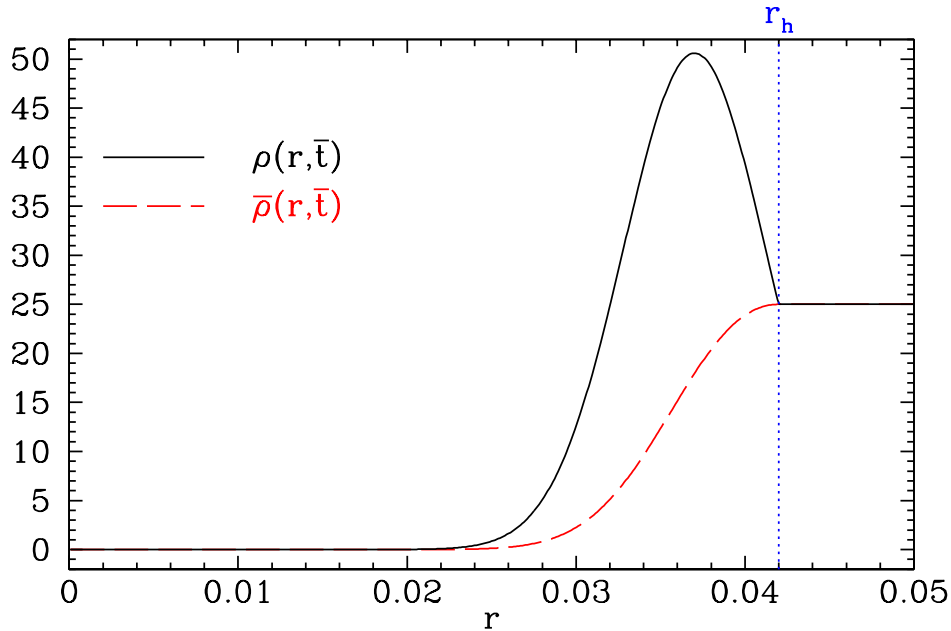}
\caption{The densities $\rho(r,\bar{t})$ (solid curve) and
$\bar{\rho}(r,\bar{t})$  (dashed curve). Here, $\bar{t}=-0.8$ (recall
$t_{BB}=-1$). The hole ends at $r_{h}=0.042$. The matching to the FRW solution
is achieved  as one can see from the plot of $\bar{\rho}(r,\bar{t})$.}
\label{rho0}
\end{center}
\end{figure}

\begin{figure}
\begin{center}
\includegraphics[width=10.5cm]{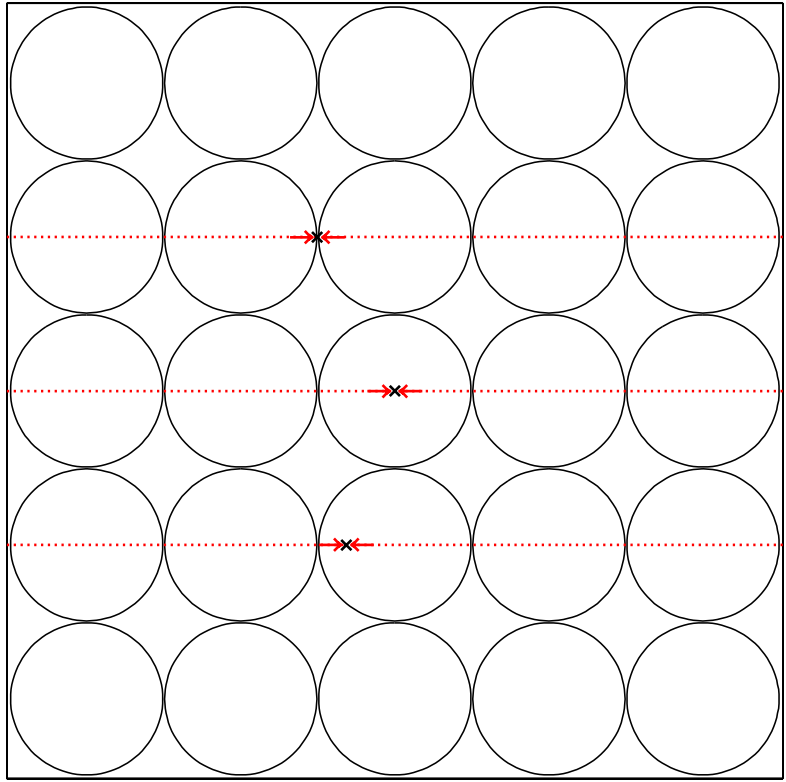}
\caption{Sketch of our swiss-cheese model. An observer at rest with 
respect to a comoving cheese-FRW observer will see an isotropic CMB 
along the two directions of sight marked with dotted red lines. 
Three possible positions for an observer are shown.}
\label{imodel}
\end{center}
\end{figure}

To be specific, we choose $\rho(r,\bar{t})$
to be
\begin{equation}
\begin{array}{ll}
\rho(r,\bar{t}) = A \exp[-(r-r_M)^2/2\sigma^2] + \epsilon \quad & (r<r_h) \\
\rho(r,\bar{t}) = \rho_{FRW}(\bar{t}) & (r > r_h),
\end{array}
\end{equation}
where $\epsilon = 0.0025$, $r_h=0.42$, $\sigma=r_h/10$, $r_M=0.037$, $A=50.59$,
and $\rho_{FRW}(\bar{t})=25$.  In Fig.\ \ref{rho0} we plot this chosen Gaussian
density profile.  The hole ends at $r_{h}=0.042$ which is\footnote{To get this number from Table \ref{units} you need to multiply $r_{h}$ by $a(t_{0})\simeq 2.92$.} $350$ Mpc and roughly
$25$ times smaller than $r_{BB}$. Note that this is not a very big bubble. But
it is an almost empty region: in the interior the matter density is roughly
$10^4$ times smaller than in the cheese. Our model consists of a sequence of up
to five holes and the observer is looking through them. The idea, however, is
that the universe is completely filled with  these holes, which form a sort of
lattice as shown in Fig.\ \ref{imodel}. In this way an observer at rest  with
respect to a comoving cheese-FRW observer will see an isotropic  CMB along the
two directions of sight shown in Fig.\ \ref{imodel}.

\begin{figure}
\begin{center}
\includegraphics[width=14 cm]{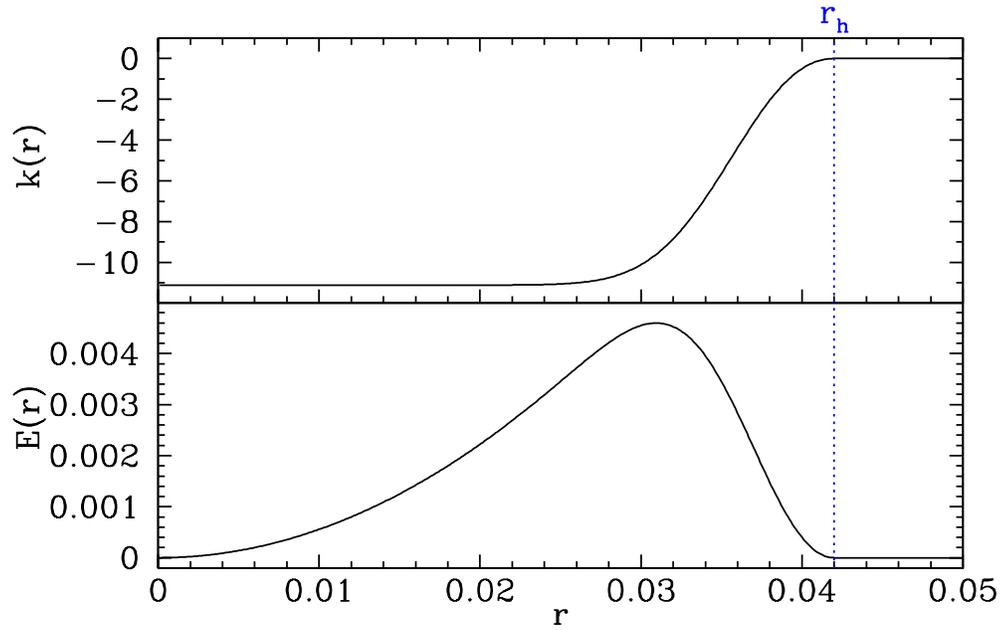}
\caption{Curvature $E(r)$ and $k(r)$ necessary for the initial conditions of no
peculiar velocities.}
\label{E}
\end{center}
\end{figure}

It is useful to consider the velocity of a shell relative to 
the FRW background.  
We define
\begin{equation}
\Delta v_{sh}(r,t)= \dot{a}_{LTB}(r,t)-\dot{a}_{FRW}(t),
\end{equation}
where $a_{LTB}(r,t)= Y(r,t)/ r$.
To have a realistic evolution, we demand that there are no initial peculiar
velocities at time $\bar{t}$, that is, to have an initial expansion $H$
independent of $r$: $\Delta v_{sh}(r,\bar{t})=0$. From Eq.\ (\ref{cucu})
this implies
\begin{equation} 
\label{Er}
E(r)=\frac{1}{2} H_{FRW}^2(\bar{t}) \, r^{2}-\frac{1}{6\pi}\frac{M(r)}{r}.
\end{equation}
The graph of $E(r)$ chosen in this way is shown in Fig.\ \ref{E}. As seen from
the figure, the curvature $E(r)$ is small compared with unity. Indeed, in many
formulae $W=(1+ 2 E)^{1/2}\simeq 1+E$ appears, therefore one should compare $E$
with $1$. In spite of its smallness, the curvature will play a crucial role to
allow a realistic evolution of structures, as we will see in the next section.

Also in Fig.\ \ref{E} we graph $k(r)=-2 E(r)/r^{2}$, which is the
generalization of the factor $k$ in the usual FRW models. (It is not normalized
to unity.) As one can see, $k(r)$ is very nearly constant  in the empty region
inside the hole. This is another way to see the reason for our choice of the
curvature function: we want to have in the center an empty bubble dominated by
negative curvature.

It is important to note that the dynamics of the hole is scale-independent:
small holes will evolve in the same way as big holes. To show this, we just
have to express Eq.\ (\ref{motion}) with respect to a generic variable
$\tilde{r}=r/g$ where $g$ fixes the scale. If we change $g$, \textit{i.e.,}
scale the density profile, we will find the same scaled shape for $k(r)$ and
the same time evolution. This property is again due to spherical symmetry
which frees the inner shells from the influence of the outer ones: We can think
of a shell as an infinitesimal FRW solution and its behavior is 
scale independent because it is a homogeneous and isotropic solution.

\clearpage
\section{The dynamics} \label{dynamics}

Now we explore the dynamics of this swiss-cheese model.  As we have said, the
cheese evolves as in the standard FRW model. Of course, inside the holes the
evolution is different. This will become clear from the plots given below. 

We will discuss two illustrative cases: a flat case where $E(r)=0$, and a
curved case where $E(r)$ is given by Eq.\ (\ref{Er}). We are really interested
only in the second case because the first will turn out to be unrealistic.  But
the flat case is useful to understand the dynamics.

\subsection{The flat case}

\begin{figure}
\begin{center}
\includegraphics[width=17 cm]{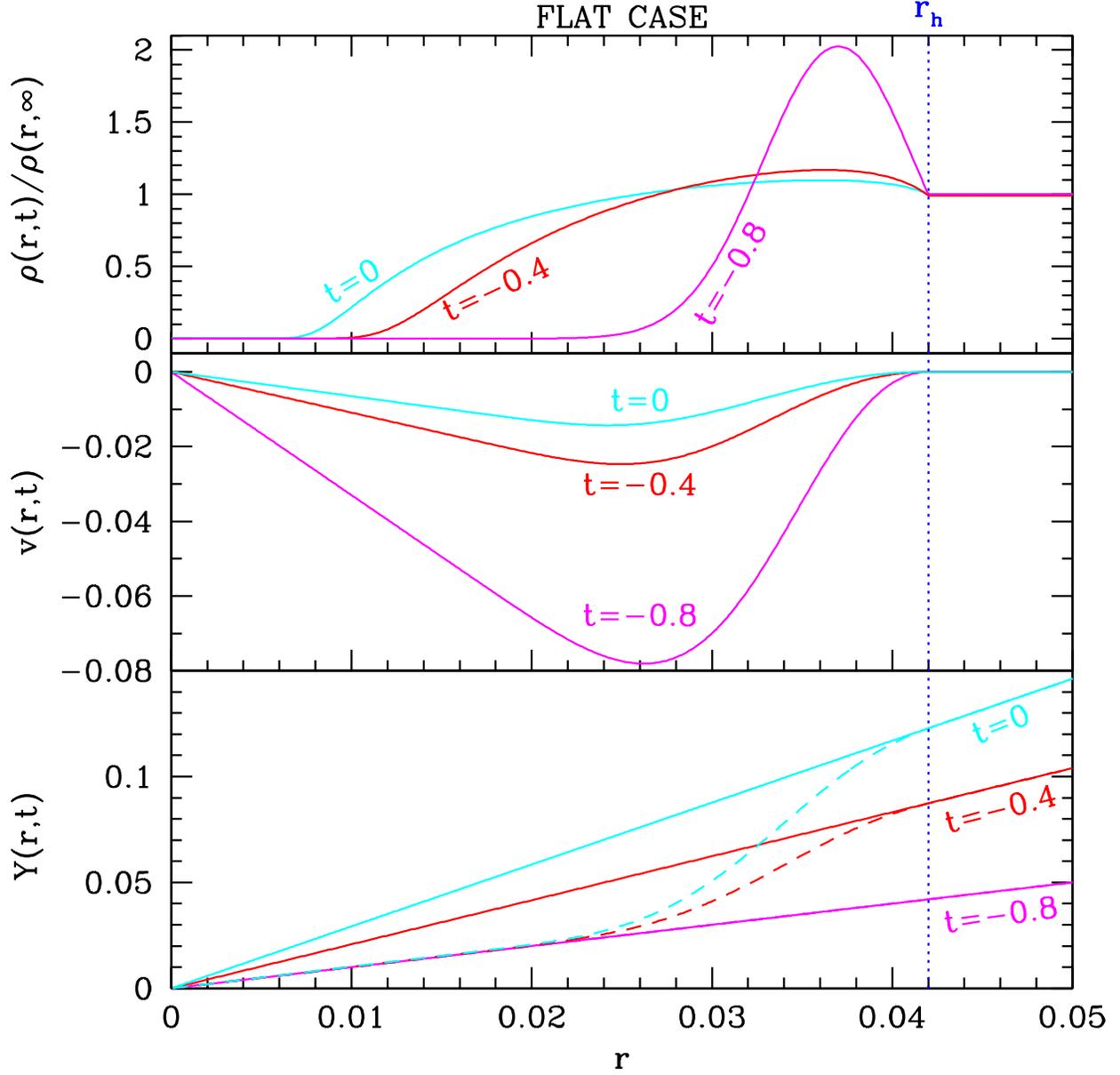}
\caption{Behavior of $Y(r,t)$ with respect to $r$, the peculiar velocities $v(r,t)$ with respect to $r$, and the
density profiles $\rho(r,t)$ with respect to $r_{FRW}=Y(r,t)/a(t)$,
for the flat case at times $t=\bar{t}=-0.8$, 
$t=-0.4$ and $t=t_0=0$.
The straight lines for $Y(r,t)$ are the FRW solutions
while the dashed lines are the LTB solutions.  For the peculiar velocities,
matter is  escaping from high density regions. The center has no peculiar
velocity  because of spherical symmetry,  and the maximum of negative peculiar
velocity is before the peak in density.  Finally, the values of $\rho(\infty,t)$ are $1,\ 2.8,$ 
and $25$, for $t=0,\ -0.4,\ -0.8$,  respectively.}
\label{flat}
\end{center}
\end{figure}

In Fig.\ \ref{flat} we show the evolution of $Y(r,t)$ for the flat case,
$E(r)=0$.  In the figure $Y(r,t)$ is plotted for three times: $t=\bar{t}=-0.8$
(recall $t_{BB}=-1$), $t = -0.4$, and $t=0$ (corresponding to today).

{}From Fig.\ \ref{flat} it is clear that outside the hole, \textit{i.e.,} for
$r \geq r_{h}$, $Y(r,t)$ evolves as a FRW solution, $Y(r,t)\propto r$. However,
deep inside the hole  where it is almost empty, there is no time evolution to
$Y(r,t)$:  it is Minkowski space.   Indeed, thanks to spherical symmetry, the
outer shells do not influence the interior. If we place additional matter
inside the empty space, it will start expanding as an FRW universe, but at a
lower rate because of the lower density. It is interesting to point out that a
photon passing the empty region will undergo no redshift: again, it is just
Minkowski space.

This counterintuitive behavior (empty regions expanding slowly) is due to the
fact that the spatial curvature vanishes.  This corresponds to an unrealistic
choice of initial peculiar velocities. To see this we plot the peculiar
velocity that an observer following a shell $r$ has with respect to an FRW
observer passing through that same spatial point. The result is also shown in
Fig.\ \ref{flat} where it is seen that matter is escaping from the high density
regions. This causes the evolution to be reversed as one can see in Fig.\
\ref{flat} from the density profile at different times: structures are not
forming, but spreading out.

Remember that $r$ is only a label for the shell
whose Euclidean position at time $t$ is $Y(r,t)$. 
In the plots of the energy density we have
normalized $Y(r,t)$ using $r_{FRW}=Y(r,t)/a(t)$.

\subsection{The curved case}

Now we move to a more interesting and relevant case. We are going to use the
$E(r)$ given by Eq.\ (\ref{Er}); the other parameters will stay the same.
Comparison with the flat case is useful to understand how the model behaves,
and in particular the role of the curvature.

\begin{figure}
\begin{center}
\includegraphics[width=17 cm]{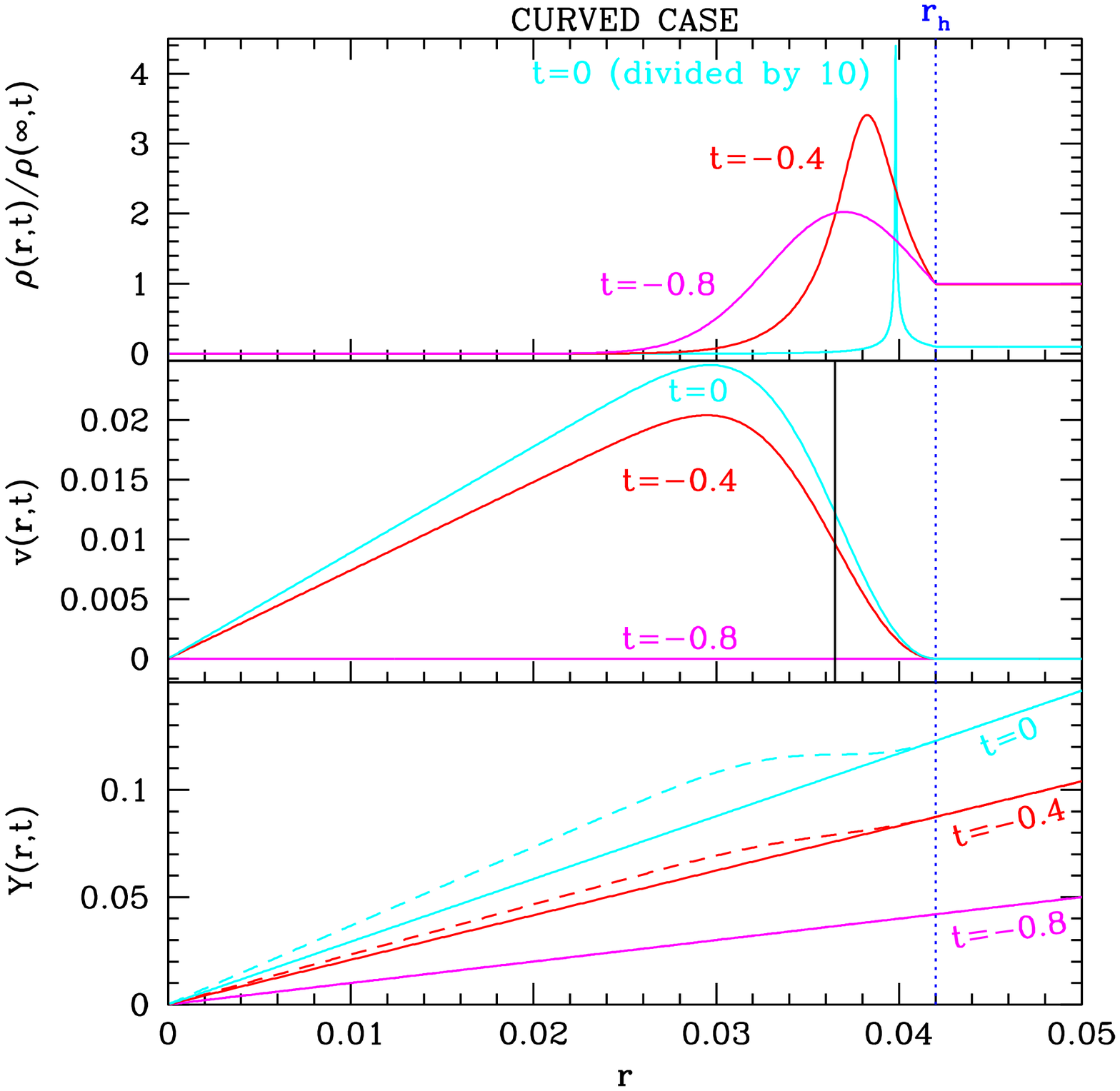}
\caption{Behavior of $Y(r,t)$ with respect to $r$, 
the peculiar velocities $v(r,t)$ with respect to $r$, and the
density profiles $\rho(r,t)$ with respect to $r_{FRW}=Y(r,t)/a(t)$,
for the curved case at times $t=\bar{t}=-0.8$, 
$t=-0.4$ and $t=t_0=0$. The straight lines for $Y(r,t)$ are the FRW solutions
while the dashed lines are the LTB solutions.  For the peculiar velocities,
the matter gradually starts to move toward high density regions. The solid
vertical line marks the position of the peak in the density with respect 
to $r$. For the
densities, note that the curve for $\rho(r,0)$ has been divided by $10$.  
Finally, the values of $\rho(\infty,t)$ are $1,\ 2.8,$ and $25$, for $t=0,\
-0.4,\ -0.8$,  respectively. }
\label{curved}
\end{center}
\end{figure}

In Fig.\ \ref{curved} the results for $Y(r,t)$ in the curved case are plotted.
Again time goes from $t=\bar{t}=-0.8$ to $t=0$ (recall that $t_{BB}=-1$ and
$t=0$ is today).

\begin{figure}
\begin{center}
\includegraphics[width=14 cm]{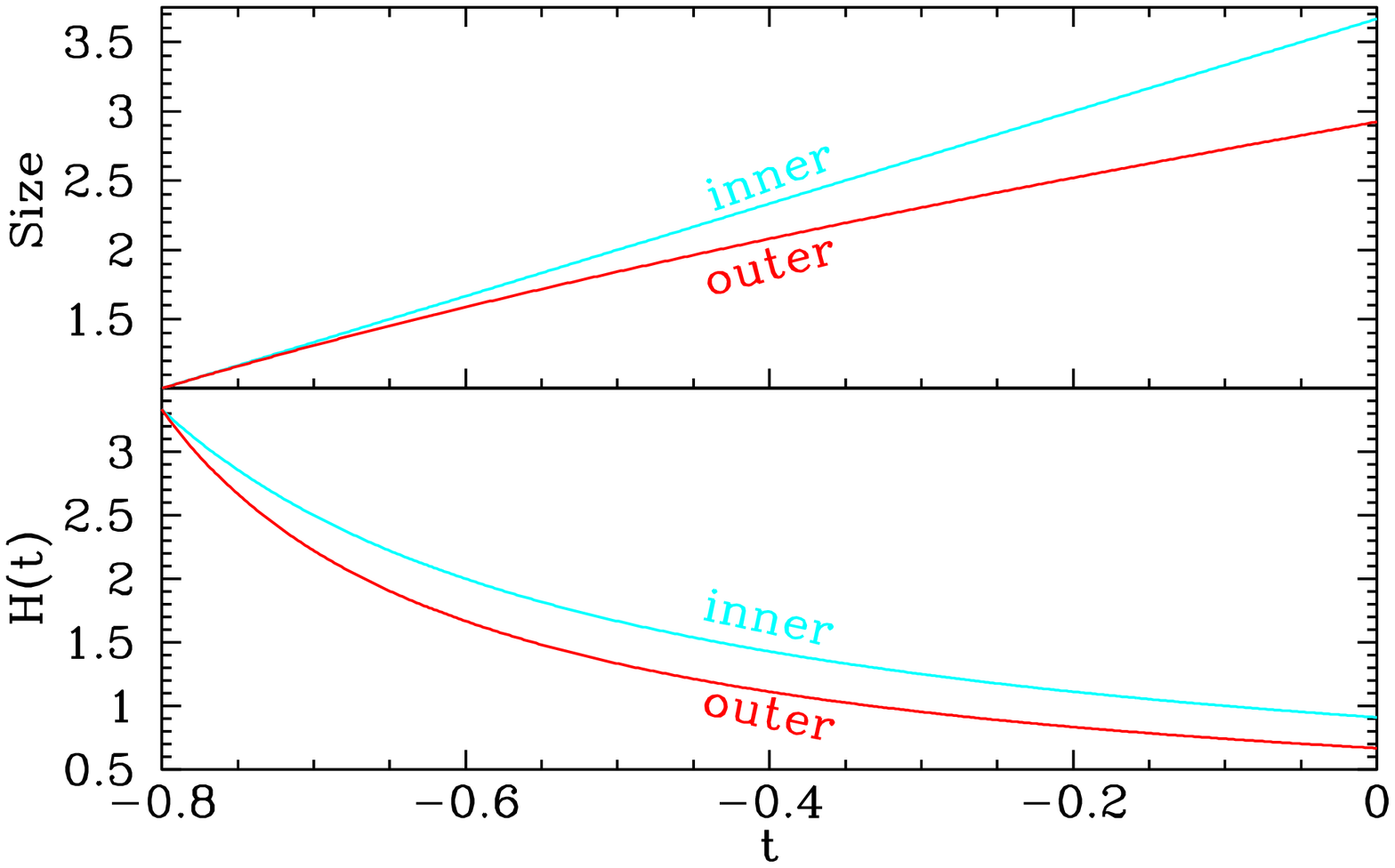}
\caption{Evolution of the expansion rate and the size for the inner 
and outer regions. Here ``inner'' refers to a point deep inside the hole, and
``outer'' refers to a point in the cheese.}
\label{outinc}
\end{center}
\end{figure}

As one can see, now the inner almost empty region is expanding faster than the
outer (cheese) region. This is shown clearly in Fig.\ \ref{outinc}, where also
the evolution of the inner and outer sizes is shown. Now the density ratio
between the cheese and the interior region of the hole increases by a factor of
$2$ between $t=\bar{t}$ and $t=0$. Initially the density ratio was $10^{4}$,
but the model is not sensitive to this number since the evolution in the
interior region is dominated by the curvature ($k(r)$ is much larger than the
matter density).

The peculiar velocities are now natural: as can be seen from Fig.\
\ref{curved}, matter is falling towards the peak in the density. The evolution
is now realistic, as one can see from Fig.\ \ref{curved}, which shows the
density profile at different times. Overdense regions start contracting and
they become thin shells (mimicking structures), while underdense regions become
larger (mimicking voids), and eventually they occupy most of the volume.

Let us explain why the high density shell forms and the nature of the shell
crossing. Because of the distribution of matter, the inner part of the hole is
expanding faster than the cheese; between these two regions  there is the
initial overdensity. It is because of this that there is less matter in the
interior part. (Remember that we matched the FRW density at the end of the
hole.) Now we clearly see what is happening: the overdense region is squeezed
by the interior and exterior regions which act as a clamp. Shell crossing
eventually happens when more shells---each labeled by its own $r$---are so
squeezed that they occupy the same physical position $Y$, that is when $Y'=0$.
Nothing happens to the photons other than passing through more shells at the
same time: this is the meaning of the $g_{r r}$ metric coefficient going to
zero.

A remark is in order here: In the inner part of the hole there is almost no
matter, it is empty. Therefore it has only negative curvature, which is largely
dominant over the matter: it is close to a Milne universe.

\newpage
\ 
\newpage
\ 
\newpage
\section{Photons} \label{photons}

\begin{figure}
\begin{center}
\includegraphics[width=14 cm]{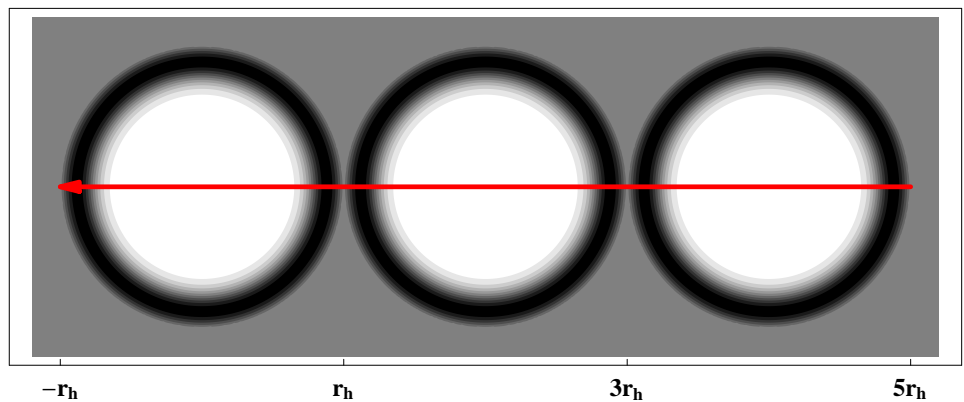}
\caption{Sketch of our model in comoving coordinates. The shading mimics 
the initial density profile: darker shading implies larger denser. The uniform 
gray is the FRW cheese. The photons pass through the holes as shown by the 
arrow.}
\label{schizzo}
\end{center}
\end{figure}

We are mostly interested in observables associated with the  propagation of
photons in our swiss-cheese model: indeed, our aim is to calculate the
luminosity-distance--redshift relation $d_{L}(z)$ in order to understand the
effects of inhomogeneities on observables. Our setup is illustrated in Fig.\
\ref{schizzo}, where there is a sketch of the model with only $3$ holes for the
sake of clarity. Notice that photons are propagating through the centers.

We will discuss two categories of cases: 1) when the observer is just outside
the last hole as in Fig.\ \ref{schizzo}, and 2) when the observer is inside the
hole. The observer in the hole will have two subcases: a) the observer located
on a high-density shell, and b) the observer in the center of the hole.  We are
mostly interested in the first case: the observer is still a usual FRW
observer, but looking through the holes in the swiss cheese.

\clearpage
\subsection{Finding the photon path: an observer in the cheese} \label{ciccio}

We will discuss now the equations we will use to find the path of a photon
through the swiss cheese. The geodesic equations can be reduced to a set of
four first-order differential equations (we are in the plane $\theta=\pi /2$):
\begin{equation}
\begin{array}{lll}
\Frac{dz}{d\lambda} & = -\Frac{\dot{Y}'}{Y'}\left((z+1)^2-
\Frac{c_\phi^2 }{Y^{2}}\right)
- c_{\phi}^{2} \Frac{\dot{Y}}{Y^{3}} & \qquad \qquad z(0)=0 \\
\Frac{dt}{d\lambda} & = z+1 & \qquad \qquad t(0)=t_{0}=0 \\
\Frac{dr}{d\lambda} & = \Frac{W}{Y'}\sqrt{(z+1)^2-\Frac{c_{\phi}^{2}}{Y^{2}}} 
 & \qquad \qquad r(0)=r_{h} \label{rgeo} \\
\Frac{d\phi}{d\lambda} & = \Frac{c_{\phi}}{Y^{2}} &  \qquad \qquad \phi(0)= \pi
\end{array}
\end{equation}
where $\lambda$ is an affine parameter that grows with time. The third
equation is actually the null condition for the geodesic. Thanks to the initial
conditions chosen we have $z(0)=0$. These equations describe the general 
path of a photon. To solve the equations we need to specify the constant 
$c_{\phi}$, a sort of angular momentum density. A first observation is 
that setting
$c_{\phi}=0$ allows us to recover the equations that describe a photon passing
radially trough the centers: $dt/dr=Y'/ W$. 

\begin{figure}
\begin{center}
\includegraphics[width=9 cm]{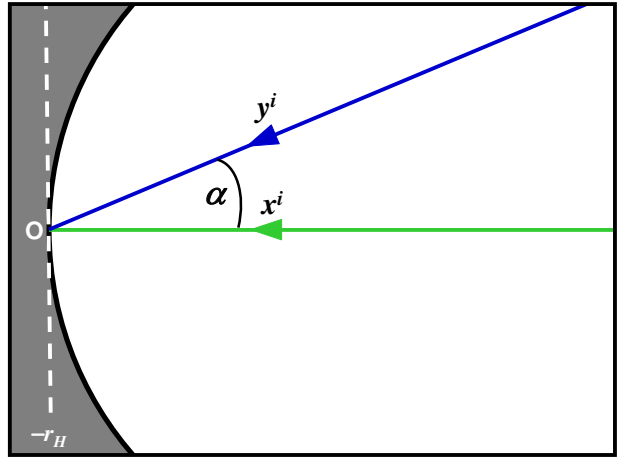}
\caption{A photon hitting the observer at an angle $\alpha$.}
\label{frecce}
\end{center}
\end{figure}

We are interested in photons that hit the observer at an angle $\alpha$ and are
passing trough all the holes as shown in Fig.\ \ref{schizzo}. To do this we
must compute the inner product of $x^{i}$ and $y^{i}$, which are the normalized
spatial vectors tangent to the radial axis and the geodesic as shown in Fig.\
\ref{frecce}. A similar approach was used in Ref.\ \cite{alnes0607}.

The inner product of $x^i$ and $y^i$ is expressed through
\begin{eqnarray}
x^{i} & = & -\frac{W}{Y'}\; (1,0,0)\vert_{\lambda=0} \\
y^{i} & = & \frac{1}{dt/d\lambda}\left. \left (\frac{d}{d\lambda},0,
\frac{d\phi}{d\lambda}\right) \right|_{\lambda=0} = \left. 
\left (\frac{dr}{d\lambda},0,\frac{d\phi}{d\lambda}\right) \right|_{\lambda=0} 
\\
x^{i}\, y^{i}\, g_{i \, j} & = & \left. \frac{Y'}{W} \; 
\frac{dr}{d\lambda}\right|_{\lambda=0}=\cos \alpha\\
c_{\phi} & = & \left. Y \sin \alpha \right| _{\lambda=0}  .
\end{eqnarray}
The vectors are anchored to the shell labeled by the value of the affine
parameter $\lambda=0$, that is, to the border of the hole. Therefore, they are
relative to the comoving observer located there. In the second equation we have
used the initial conditions given in the previous set of equations, while to
find the last equation we have used the null condition evaluated at
$\lambda=0$. 

The above calculations use coordinates relative to the center. However, the
angle $\alpha$ is a scalar in the hypersurface we have chosen: we are using the
synchronous and comoving gauge. Therefore, $\alpha$ is the same angle 
measured by a comoving observer of Fig.\ \ref{frecce} located on the shell 
$r=-r_{h}$: it is a coordinate transformation within the same hypersurface.

Given an angle $\alpha$ we can solve the equations. We have to change the sign
in Eq.\ (\ref{rgeo}) when the photon is approaching the center with respect to
the previous case where it is moving away. Also, we have to sew together the
solutions between one hole and another, giving not only the right initial 
conditions, but also the appropriate constants $c_{\phi}$  (see Appendix
\ref{sewing}).

Eventually we end up with the solution $t(\lambda)$, $r(\lambda)$,
$\phi(\lambda)$ and $z(\lambda)$ from which we can calculate the observables of
interest.

\subsection{Finding the photon path: an observer in the hole}

Finding the solution in this case is the same as in the previous case with the
only difference that in Eq.\ (\ref{rgeo}) the initial condition is now
$r(0)=r_{obs}$. But this observer has a peculiar velocity with respect to an
FRW observer passing by. This, for example, will make the observer see an
anisotropic cosmic microwave background as it is clear from Fig.\ \ref{imodel}.
This Doppler effect, however, is already corrected in the solution we are going
to find since we have chosen $z(0)=0$ as initial condition.

\begin{figure}
\begin{center}
\includegraphics[width=9 cm]{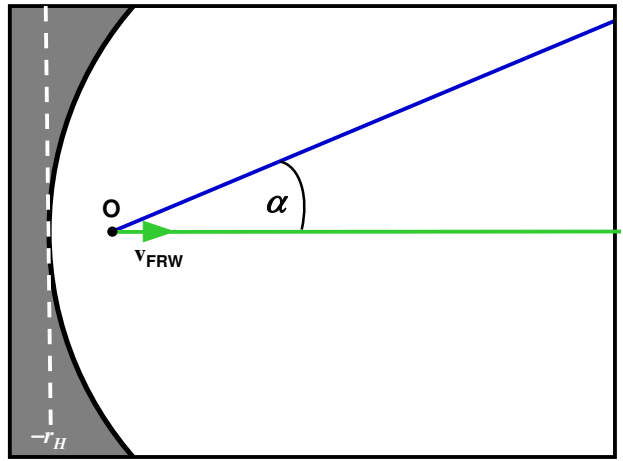}
\caption{A comoving observer and a FRW observer live in different frames, this
results in a relative velocity $v_{FRW}$ between observers.}
\label{doppler}
\end{center}
\end{figure}

There is however also the effect of light aberration which changes the angle
$\alpha$ seen by the comoving observer with respect to the angle
$\alpha_{\scriptscriptstyle FRW}$ seen by an FRW observer. The photon can be
thought as coming from a source very close to the comoving observer: therefore
there is no peculiar motion between them.  The FRW observer is instead moving
with respect to this reference frame as pictured in Fig.\ \ref{doppler}. The
relation between $\alpha$ and $\alpha_{\scriptscriptstyle FRW}$ is given by the
relativistic aberration formula:
\begin{equation}
\cos \alpha_{\scriptscriptstyle FRW}=\frac{\cos \alpha+ v/c}{1+v/c 
\; \cos \alpha} .
\end{equation}
The angle changes because the hypersurface has been changed. The velocity will
be taken from the calculation (see Fig.\ \ref{curved} for the magnitude of the
effect).

\subsection{Distances}

The angular diameter distance is defined as:
\begin{equation}
d_{A}=\frac{D}{\alpha_{\scriptscriptstyle FRW}} ,
\end{equation}
where $D$ is the proper diameter of the source and $\alpha$ is the angle at
which the source is seen by the observer. Using this definition to find $d_{A}$
we have
\begin{equation}
d_A = \frac{2 \, Y(r(\lambda),t(\lambda)) \; \sin \phi(\lambda)}{2 \,  
\alpha_{\scriptscriptstyle FRW}}  .
\end{equation}
The luminosity distance will then be:
\begin{equation}
d_{L}=(1+z)^{2} d_{A} .
\end{equation}
The formula we are going to use for $d_{A}$ is exact in the limit of zero
curvature. However in our model $E(r)$ is on average less than $0.3\%$  and
never more than $0.4\%$, as it can be seen from Fig.\ \ref{E}:  therefore the
approximation is good. Moreover, we are interested mainly in the case when the
source is out of the last hole as pictured in Fig.\ \ref{schizzo}, and in this
case the curvature is exactly zero and the result is exact.

We have checked that the computation of $d_{A}$ is independent of $\alpha$ for
small angles and that the result using the usual FRW equation coincides with
theoretical prediction for $d_{A}$. We also checked that $d_{A}$ reduces to
$Y(r,t)$ when the observer is in the center.

Finally we checked our procedure in comparison with the formula ($E.31$) of
Ref.\ \cite{notari}: this is a rather different way to find the angular
distance and therefore this agreement serves as a consistency check. We placed
the observer in the same way and we found the same results provided that we use
the angle $\alpha$ uncorrected for the light-aberration effect.

\section{Results: observer in the cheese} \label{cheese}

Now we will look through the swiss cheese comparing the results with respect to
a FRW-EdS universe and a $\Lambda$CDM case.

We will first analyze in detail the model with five holes, which is the one 
which we are most interested in. 
For comparison, we will study models with one big
hole and one small hole. In the model with one big hole, the hole will be
five-times bigger in size than in the model with five holes: \textit{i.e.,}
they will cover the same piece of the universe.

The observables on which we will focus are the changes in redshift
$z(\lambda)$, angular-diameter distance $d_{A}(z)$, luminosity distance
$d_{L}(z)$, and the corresponding distance modulus $\Delta m(z)$.

\subsection{Redshift histories} \label{histories}

\begin{figure}
\begin{center}
\includegraphics[width=16.2 cm]{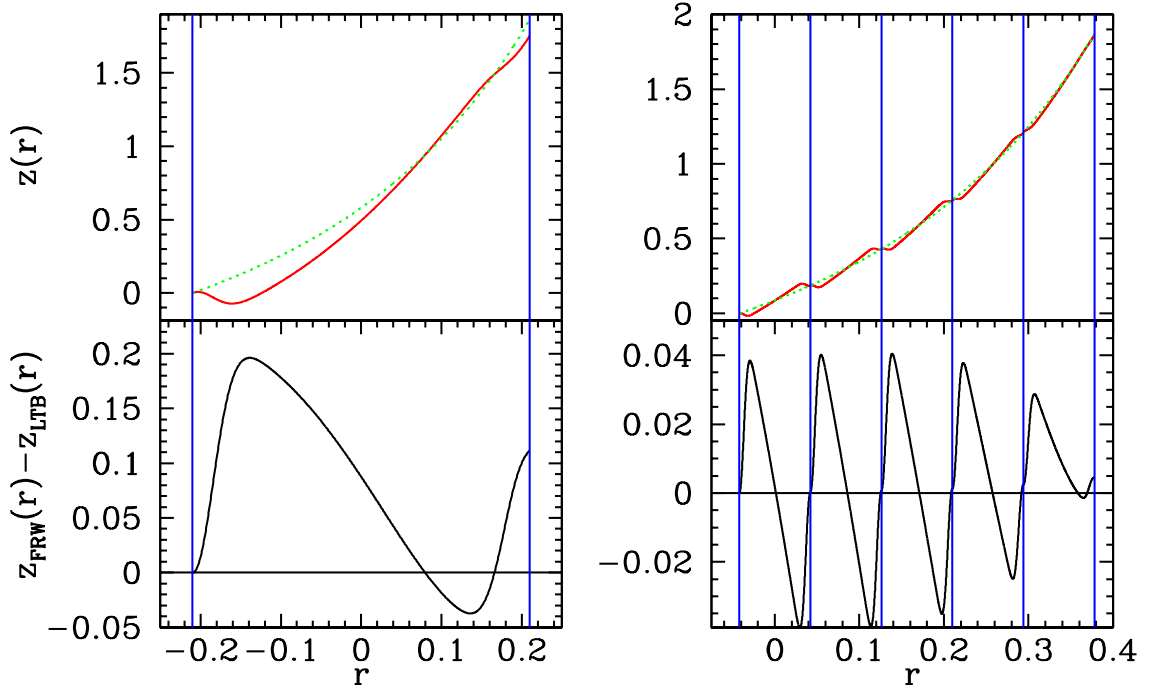}
\caption{Redshift histories for a photon that travels from one side of the 
one-hole chain (left) and five-hole chain (right) to the other where the 
observer will detect it at present time. The ``regular'' curve is for the FRW
model. The vertical lines mark the edges of the holes. The plots are with
respect to the coordinate radius $r$. Notice also that along the voids the
redshift is increasing  faster: indeed $z'(r)=H(z)$ and the voids are expanding
faster.}
\label{zorro}
\end{center}
\end{figure}

Now we will first compare the redshift undergone by photons that travel through
the model with either five holes or one hole to the FRW solution of the cheese.
In Fig.\ \ref{zorro} the results are shown for a photon passing through the
center with respect to the coordinate radius. As one can see, the effects 
of the inhomogeneities on the redshift are smaller in the five-hole case.

It is natural to expect a compensation, due to the spherical symmetry,
between the ingoing path and the outgoing one inside the same hole. This
compensation is evident in Fig.~\ref{zorro}.

However, there is a compensation already on the scale of half a hole as it is
clear from the plots. This mechanism is due to the density profile chosen, that
is one whose average matches the FRW density of the cheese: roughly speaking we
know that $z'=H \propto \rho = \rho_{\scriptscriptstyle FRW} + \delta \rho$. We
chose the density profile in order to have $\langle \delta \rho \rangle=0$, and
therefore in its journey from the center to the border of the hole the photon
will see a $\langle H\rangle \sim H_{\scriptscriptstyle FRW}$ and therefore
there will be compensation for $z'$.

Let us see this analytically. We are interested in computing a line average of
the expansion along the photon path in order to track what is going on.
Therefore, we shall not use the complete expansion scalar:
\begin{equation}
\theta=\Gamma_{0k}^{k}=2\frac{\dot{Y}}{Y}+\frac{\dot{Y}'}{Y'} ,
\end{equation}
but, instead, only the part of it pertinent to a radial line average:
\begin{equation} \label{dito}
\theta_r=\Gamma_{01}^{1}=\frac{\dot{Y}'}{Y'}\equiv H_{r} ,
\end{equation}
where $\Gamma_{0k}^{k}$ are the Christoffel symbols and $\theta$ is the trace
of the extrinsic curvature.

Using $H_r$, we obtain:
\begin{equation}
\langle H_r \rangle =
\frac{\int_{0}^{r_{h}}dr \; H_r \; Y' / W}
{\int_{0}^{r_{h}} dr \; Y' / W} \simeq \left. \frac{\dot{Y}}{Y}\right|_{r=r_{h}}
= H_{\scriptscriptstyle FRW} ,
\end{equation}
where the approximation comes from neglecting the (small) curvature and the
last equality holds thanks to the density profile chosen. This is exactly the
result we wanted to find. However, we have performed an average at constant time
and therefore we did not let the hole and its structures  evolve while the
photon is passing: this effect will partially break the compensation. This
sheds light on the fact that photon physics seems to be affected by the
evolution of the inhomogeneities more than by the inhomogeneities themselves.
We can argue that there should be perfect compensation if the hole will have a
static metric such as the Schwarzschild one. In the end, this is a limitation
of our assumption of spherical symmetry.

This compensation is almost perfect in the five-hole case, while it is not in
the one-hole case: in the latter case the evolution has more time to change the
hole while the photon is passing. Summarizing, the compensation is working on
the scale $r_{h}$ of half a hole. These results are in agreement  with Ref.\
\cite{notari07}.

\begin{figure}
\begin{center}
\includegraphics[width=13 cm]{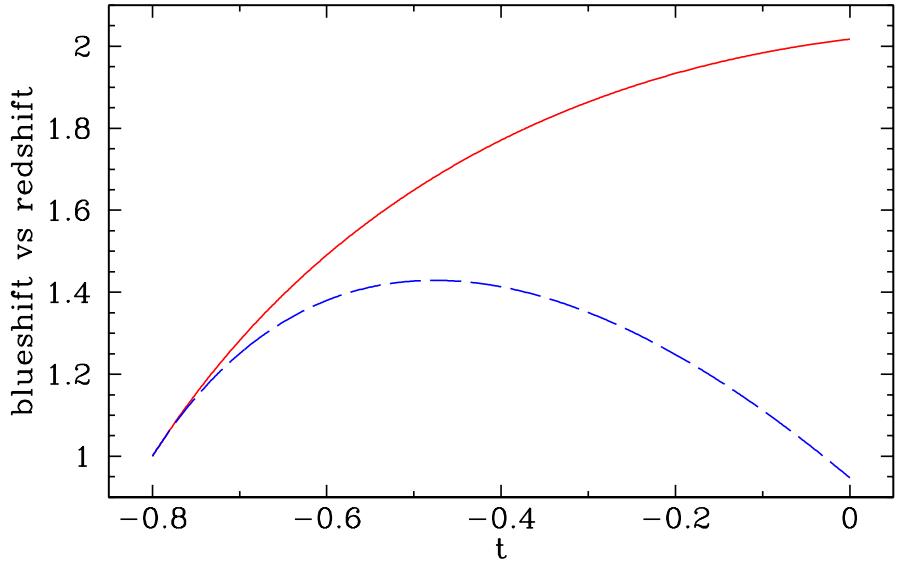}
\caption{Distance between the observer and two different shells. In the
solid curve $r=0.55 \, r_{h}$ will be redshifted, while in the dashed curve,
$r=0.8 \, r_{h}$ will be blueshifted. The latter indeed will start to collapse 
toward the observer. Time goes from $t=-0.8$ to $t=0$.
The observer is located just outside of the hole at $r=r_{h}$.}
\label{blue}
\end{center}
\end{figure}

{}From the plot of the redshift one can see that the function $z(r)$ is not
monotonic. This happens at recent times when the high-density thin shell forms.
This blueshift is due to the peculiar movement of the matter that is forming
the shell. This feature is shown in Fig.\ \ref{blue} where the distance between
the observer located just out of the hole at $r=r_{h}$ and two different shells
is plotted. In the solid curve one can see the behavior with respect to a
normal redshifted shell, while in dashed curve one can see the behavior with
respect to a shell that will be blueshifted: initially the distance increases
following the Hubble flow, but when the shell starts forming, the peculiar
motion prevails on the Hubble flow and the distance decreases during the
collapse.

It is finally interesting to interpret the redshift that a photon undergoes
passing the inner void. The small amount of matter is subdominant with respect
to the curvature which is governing the evolution, but still it is important to
define the space: in the limit of zero matter in the interior of the hole, we
recover a Milne universe, which is just (half of) Minkowski space in unusual
coordinates. Before this limit the redshift was conceptually due to the
expansion of the spacetime, after this limit it is instead due to the peculiar
motion of the shells which now carry no matter: it is a Doppler effect.

\subsection{Luminosity and Angular-Diameter Distances}

\subsubsection{The five-hole model} \label{5holes}

\begin{figure}
\begin{center}
\includegraphics[width=17 cm]{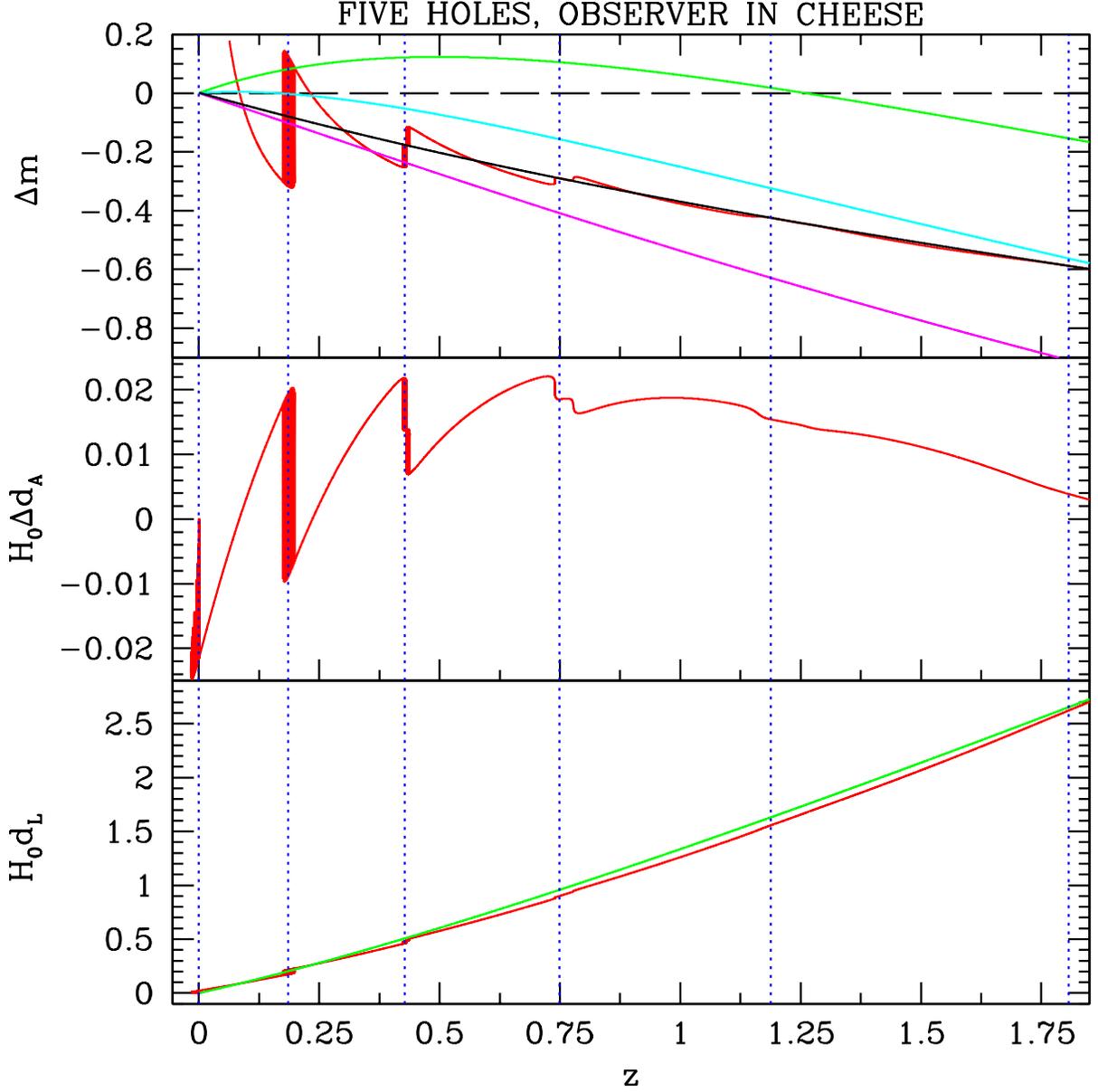}
\caption{On the bottom the luminosity distance $d_L(z)$ in the  five-hole model
(jagged curve) and the $\Lambda$CDM solution with $\Omega_{M}=0.6$ and
$\Omega_{DE}=0.4$ (regular curve) are shown.  In the middle is the change in
the angular diameter distance, $\Delta d_A(z)$, compared to a $\Lambda$CDM
model with  $\Omega_{M}=0.6$ and $\Omega_{DE}=0.4$. The top panel shows the
distance modulus in various cosmological models. The jagged line is for the
five-hole LTB model. The regular curves, from top to bottom, are a $\Lambda$CDM
model with  $\Omega_{M}=0.3$ and $\Omega_{DE}=0.7$, a $\Lambda$CDM model with  
$\Omega_{M}=0.6$ and $\Omega_{DE}=0.4$, the best smooth fit to the LTB model,
and the EdS model.  The vertical lines mark the edges of the five holes.}
\label{5incheese}
\end{center}
\end{figure}

In Fig.\ \ref{5incheese} the results for the luminosity distance and angular
distance are shown. The solution is compared to the one of the $\Lambda$CDM
model with $\Omega_{M}=0.6$ and $\Omega_{DE}=0.4$. Therefore, we have an
effective $q_{0}=\Omega_{M}/2-\Omega_{DE}=-0.1$. In all the plots we will
compare this $\Lambda$CDM solution to our swiss-cheese solution. The strange
features which appear near the contact region of the holes at recent times are
due to the non-monotonic behavior of $z(r)$, which was explained in the
previous section.

The distance modulus is plotted in the top panel of Fig.\ \ref{5incheese}. The
solution shows an oscillating behavior which is due to the simplification of
this toy model in which all the voids are concentrated inside the holes and all
the structures are in thin spherical shells. For this reason a fitting curve
was plotted: it is passing through the points of the photon path that are in
the cheese between the holes. Indeed, they are points of average behavior and
represent well the coarse graining of this oscillating curve. The
simplification of this model tells us also that the most interesting part of
the plot is farthest from the observer, let us say at $z>1$. In this region we
can see the effect of the holes clearly: they move the curve from the EdS
solution (in purple) to the $\Lambda$CDM one with $\Omega_{M}=0.6$ and
$\Omega_{DE}=0.4$ (in blue). Of course, the model in not realistic enough to
reach the ``concordance'' solution.

Here we discuss a comparison of our results with those of Ref.\
\cite{notari07}. In that paper they do not find the large difference from FRW
results that we do.  First of all, we note that we are able to reproduce their
results using our techniques.  The difference between their results and ours is
that our model has very strong nonlinear evolution, in particular, close to
shell crossing where we have to stop the calculation. The authors of Ref.\
\cite{notari07} also used smaller holes with a different 
density/initial-velocity profile.  This demonstrated that a large change in
observables may require either non-spherical inhomogeneities, or evolution very
close to shell crossing.  (We remind the reader that caustics are certainly
expected to form in cold dark matter models.)

\begin{figure}
\begin{center}
\includegraphics[width=16.2 cm]{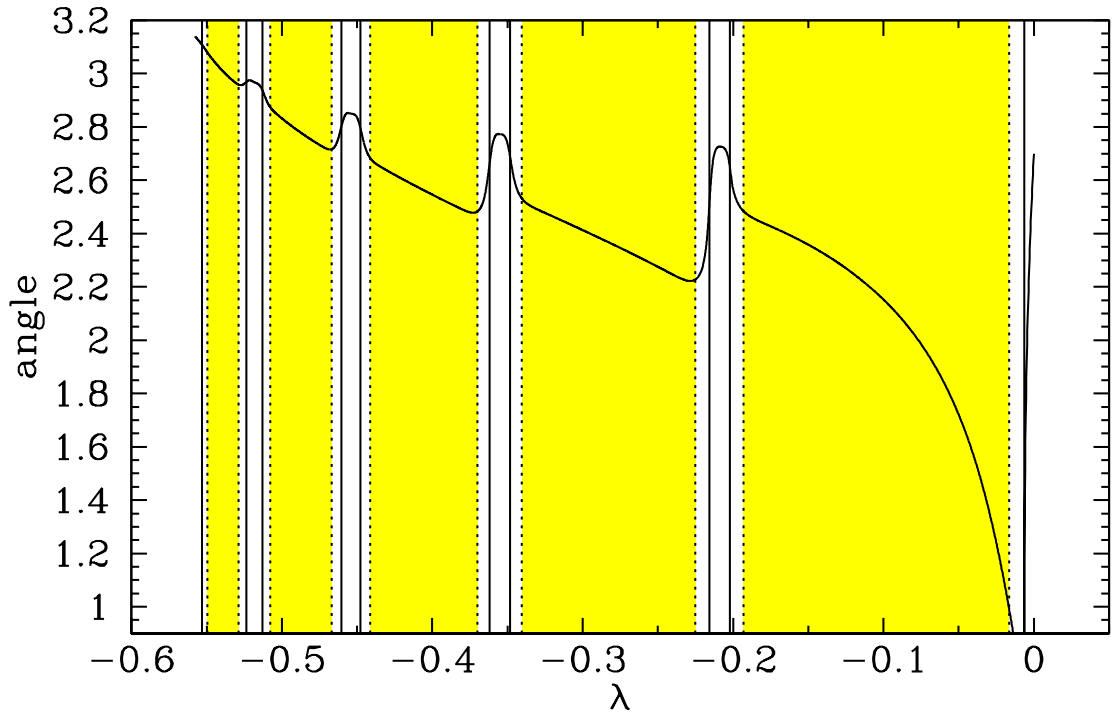}
\caption{The angle from the observer is plotted. The dashed vertical lines 
near the empty region mark the shell of maximum peculiar velocities of 
Fig.\ \ref{curved}. The shaded regions represent the inner FRW solution. The 
solid vertical lines mark the peak in density. The angle at which the photon 
hits the observer is $2.7\,^{\circ}$ on the left.}
\label{bend}
\end{center}
\end{figure}

Let us return now to the reason for our results. As we have seen previously,
due to spherical symmetry there are no significant redshift effects in the
five-hole case. Therefore, these effects must be due to changes in the
angular-diameter distance. Fig.\ \ref{bend} is useful to understand what is
going on: the angle from the observer is plotted. Through the inner void and
the cheese the photon is going straight: they are both $FRW$ solutions even if
with different parameters. This is shown in the plot by constancy of the slope.
The bending occurs near the peak in the density where the $g_{r \, r}$
coefficient of the metric goes toward zero. Indeed the coordinate velocity of
the photon can be split in an angular part: $v_{\phi}=d\phi/dt=1/\sqrt{g_{\phi
\, \phi}}$ and a radial part $v_{r}=dr/dt=1/\sqrt{g_{r \, r}}$. While
$v_{\phi}$ behaves well near the peak, $v_{r}$ goes to infinity in the limit
where shell crossing is reached: the photons are passing more and more
matter shells in a short interval of time as the evolution approaches the
shell-crossing point.  Although in our model we do not reach shell crossing,
this is the reason for the bending. We therefore see that all the effects in
this model, redshift and angular effects, are due to the evolution of the
inhomogeneities.

\subsubsection{The one-hole model: the big hole case} 

\begin{figure}
\begin{center}
\includegraphics[width=17 cm]{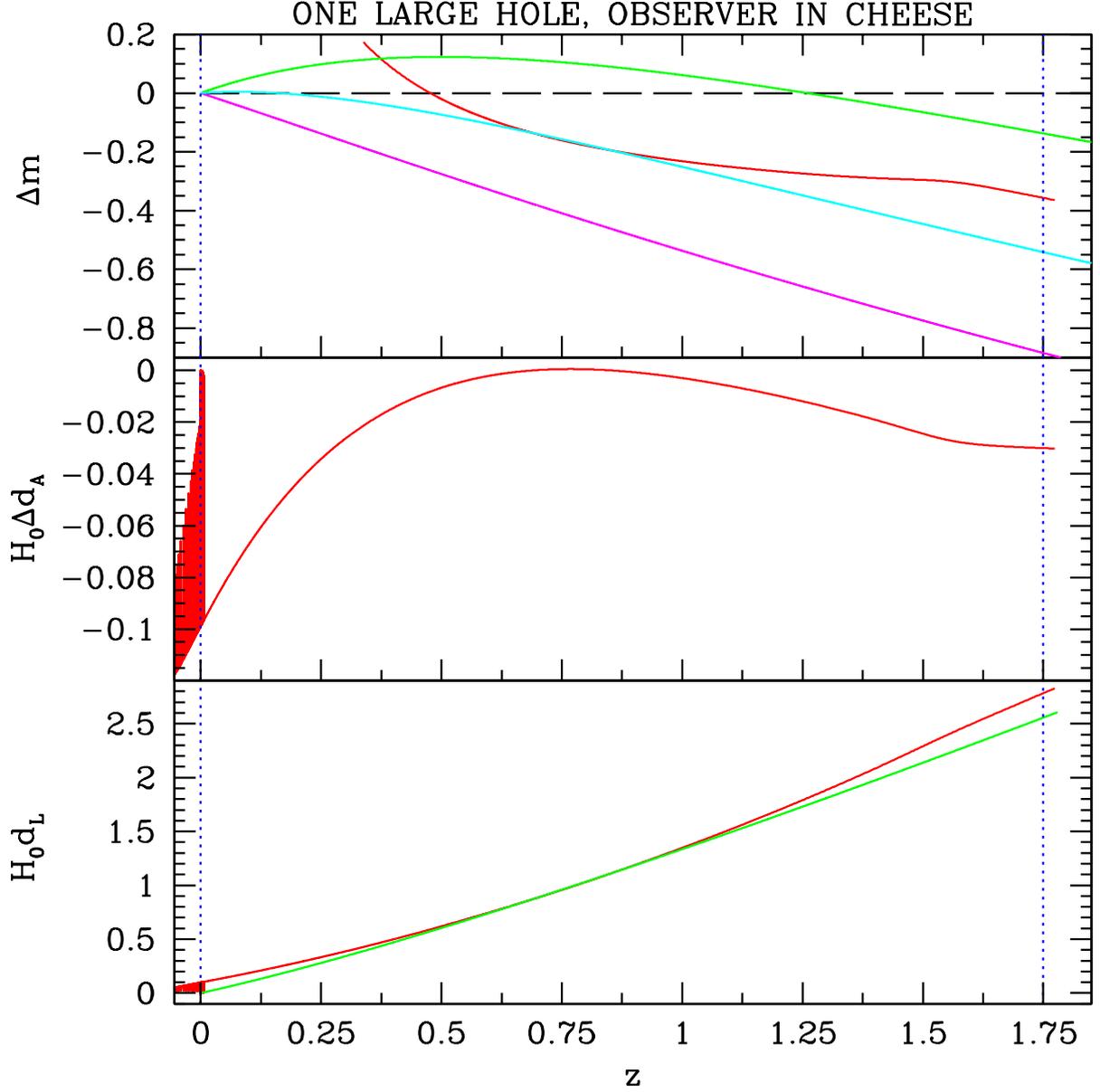}
\caption{On the bottom is shown the luminosity distance $d_L(z)$ in the one-hole
model (jagged curve) and the $\Lambda$CDM solution with $\Omega_{M}=0.6$ and
$\Omega_{DE}=0.4$ (regular curve).  In the middle is the change in the the
angular diameter distance, $\Delta d_A(z)$, compared to a $\Lambda$CDM model
with  $\Omega_{M}=0.6$ and $\Omega_{DE}=0.4$. On the top is shown the distance
modulus in various cosmological models. The jagged line is for the one-hole LTB
model.  The regular curves, from top to bottom are a $\Lambda$CDM model with 
$\Omega_{M}=0.3$ and $\Omega_{DE}=0.7$, a $\Lambda$CDM model with
$\Omega_{M}=0.6$ and $\Omega_{DE}=0.4$ and the EdS model.  The vertical lines
mark the edges of the hole.}
\label{1big}
\end{center}
\end{figure}

Let us see now how the results change if instead of the five-hole model we use
the one-hole model.  We have already shown the redshift results in the previous
section. As one can see from Fig.\ \ref{1big} the results are more dramatic:
for high redshifts the swiss-cheese curve can be fit by a $\Lambda$CDM model
with less dark energy than $\Omega_{DE}=0.6$ as in the five-hole model.
Nonetheless, the results have not changed so much compared to the change in the
redshift effects discussed in the previous section. Indeed the compensation
scale for angular effects is $2 r_{h}$ while the one for redshift effects is
$r_{h}$.

\subsubsection{The one-hole model: the small hole case}

\begin{figure}
\begin{center}
\includegraphics[width=17 cm]{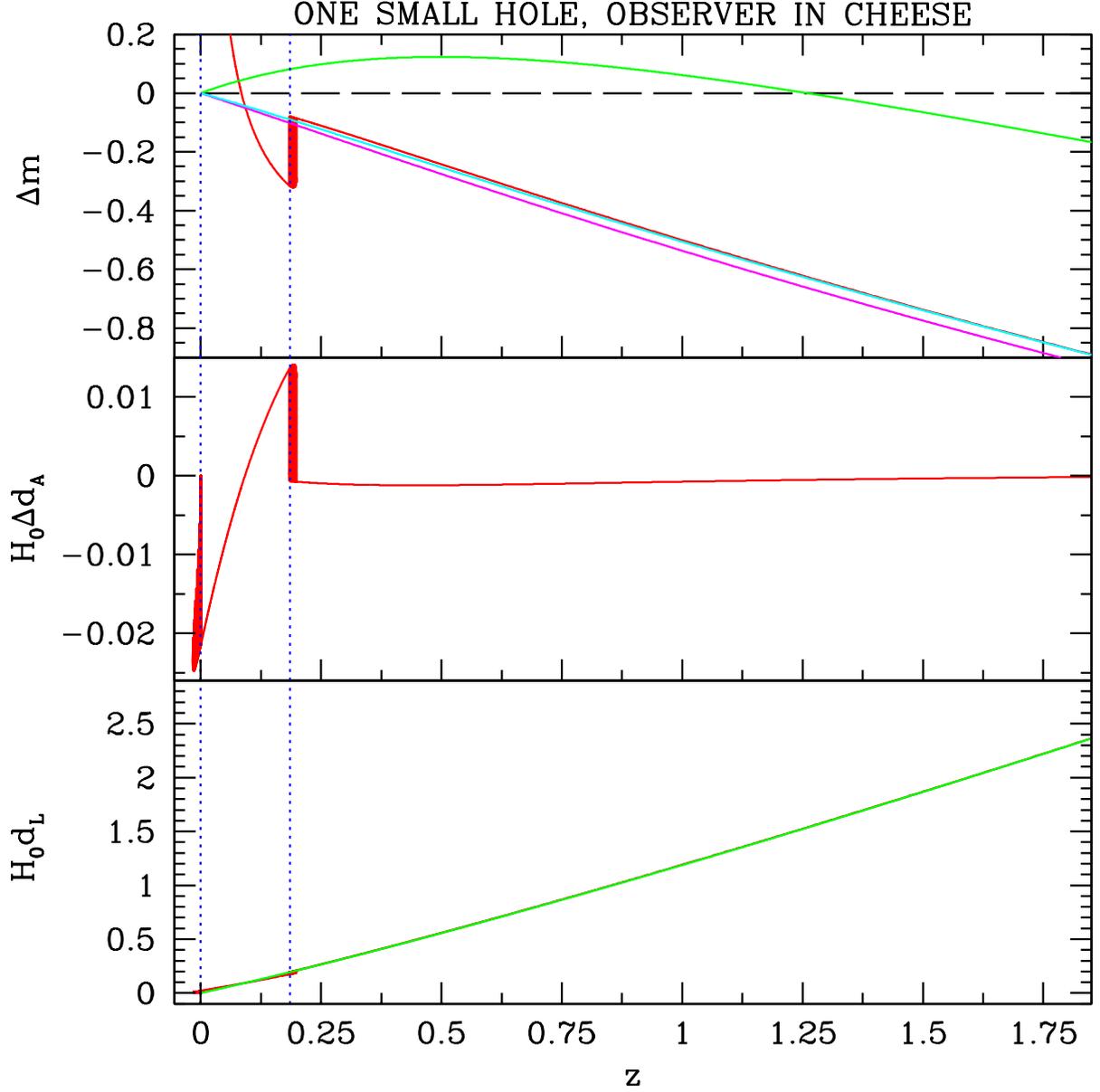}
\caption{On the bottom is shown the luminosity distance $d_L(z)$ in the 1-hole
model (jagged curve) and the $\Lambda$CDM solution with $\Omega_{M}=0.95$ and
$\Omega_{DE}=0.05$ (regular curve).  In the middle is the change in the the
angular diameter distance, $\Delta d_A(z)$, compared to a $\Lambda$CDM model
with  $\Omega_{M}=0.95$ and $\Omega_{DE}=0.05$. On the top is shown the distance
modulus in various cosmological models. The jagged line is for the one-hole LTB
model.  The regular curves, from top to bottom are a $\Lambda$CDM model with 
$\Omega_{M}=0.3$ and $\Omega_{DE}=0.7$, a $\Lambda$CDM model with
$\Omega_{M}=0.95$ and $\Omega_{DE}=0.05$ and the EdS model.  The vertical lines
mark the edges of the hole.}
\label{1small}
\end{center}
\end{figure}

Finally if we remove four holes from the five-hole model, we lose almost all
the effects. This is shown in Fig.\ \ref{1small}: now the model can be compared
to a $\Lambda$CDM model with $\Omega_{M}=0.95$ and $\Omega_{DE}=0.05$.

\clearpage
\section{Results: observer in the hole} \label{hole}

Now we will examine the case in which the observer is inside the last hole in
the five-hole model. We will first put the observer on the high-density shell
and then place the observer in the center.

\subsection{Observer on the high density shell}

\begin{figure}
\begin{center}
\includegraphics[width=15 cm]{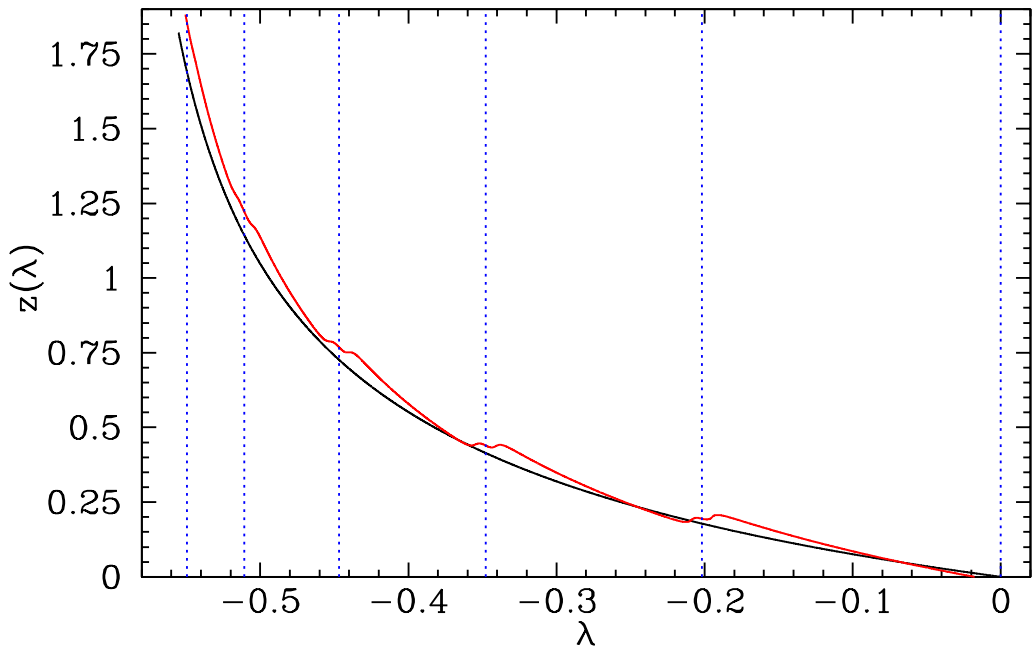}
\caption{Redshift histories for a photon that travels through the 
five-hole-chain to the observer placed on the high density shell.
The ``regular'' line is for the FRW model. $\lambda$ is the affine 
parameter and it grows with the time which go from the left to the right. 
The vertical lines mark the end and the beginning of the holes.}
\label{zpik}
\end{center}
\end{figure}

\begin{figure}
\begin{center}
\includegraphics[width=17 cm]{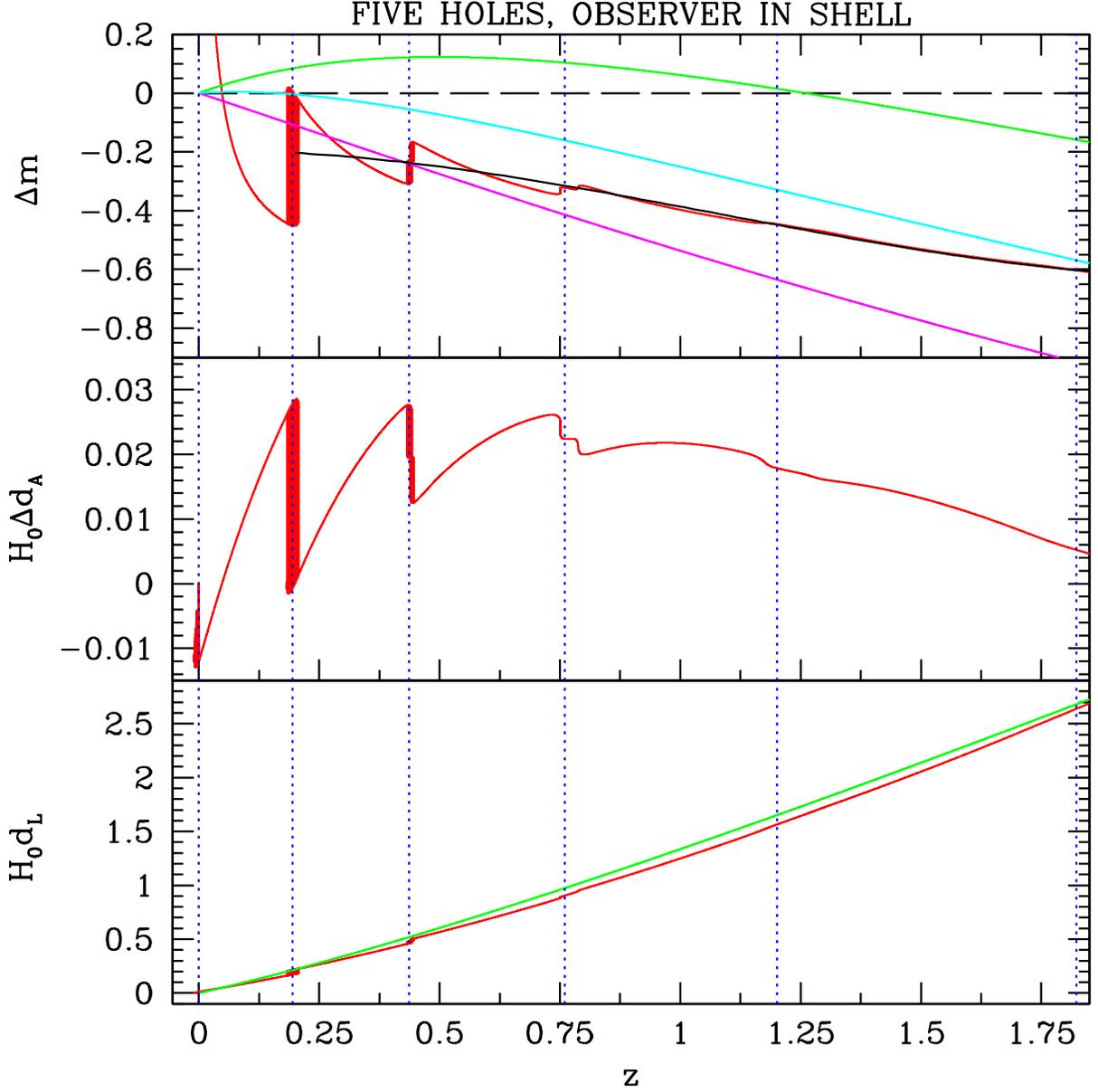}
\caption{On the bottom is shown the luminosity distance $d_L(z)$ in the
five-hole model (jagged curve) and the $\Lambda$CDM solution with
$\Omega_{M}=0.6$ and $\Omega_{DE}=0.4$ (regular curve). In the middle is the
change in the angular diameter distance, $\Delta d_A(z)$, compared to a
$\Lambda$CDM model with  $\Omega_{M}=0.6$ and $\Omega_{DE}=0.4$. On the top is
shown the distance modulus in various cosmological models. The jagged line is
for the five-hole LTB model.  The regular curves, from top to bottom are  a
$\Lambda$CDM model with  $\Omega_{M}=0.3$ and $\Omega_{DE}=0.7$, a $\Lambda$CDM
model with   $\Omega_{M}=0.6$ and $\Omega_{DE}=0.4$, the best smooth fit to the
LTB model, and the EdS model.  The vertical lines mark the edges of the five
holes.}
\label{shell}
\end{center}
\end{figure}

In the section we show the results for the observer on the high-density shell.
As one can see from Fig.\ \ref{zpik}, now the compensation in the redshift
effect is lost: the photon is not completing the entire last half of the last
hole. The results for the luminosity distance and the angular distance do not
change much as shown in Fig.\ \ref{shell}.

Remember that in this case the observer has a peculiar velocity compared to the
FRW observer passing through the same point. We correct the results taking into
account both the Doppler effect and the light aberration effect.

\subsection{Observer in the center}

\begin{figure}
\begin{center}
\includegraphics[width=15 cm]{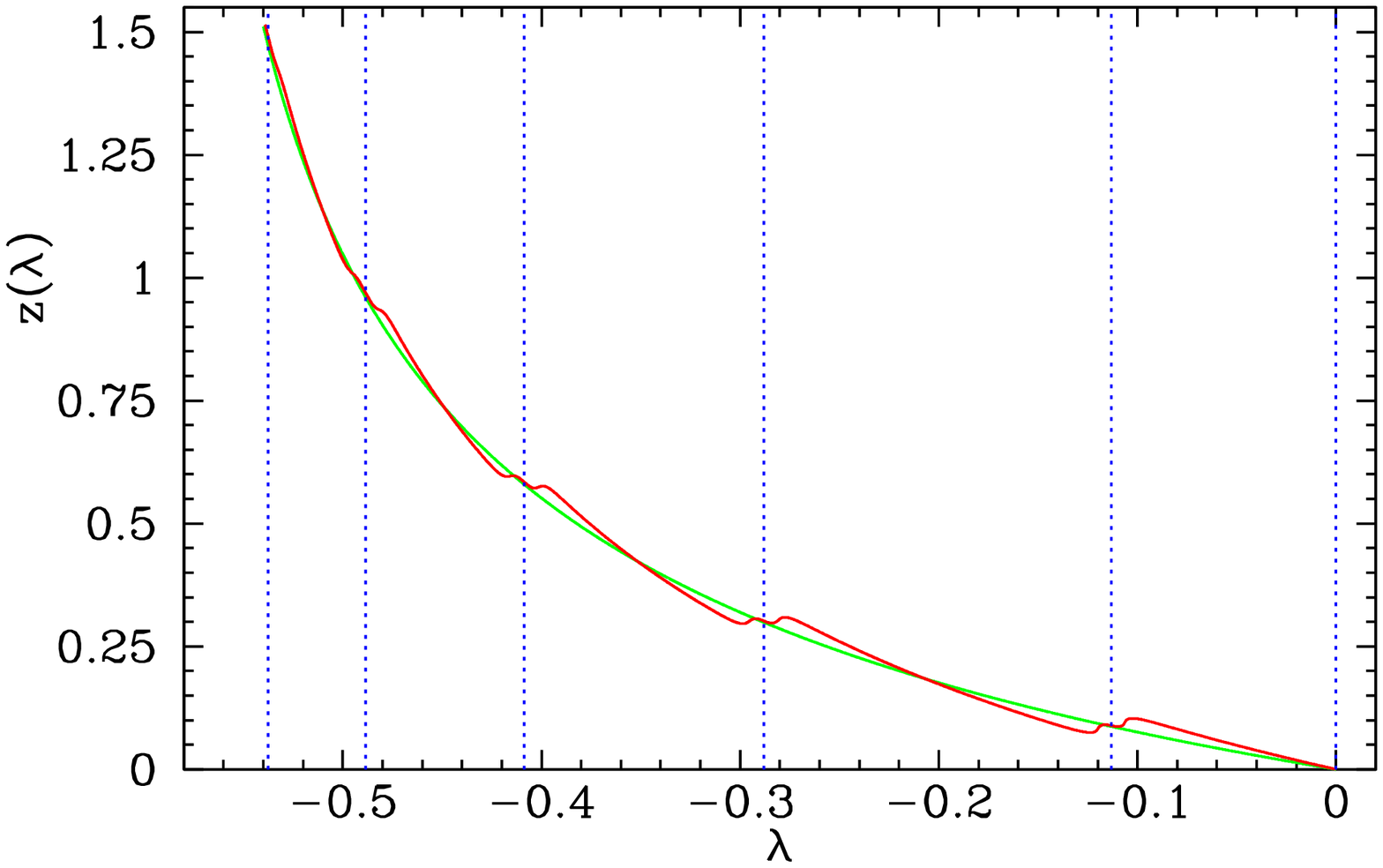}
\caption{Redshift histories for a photon that travels through the 
five-hole-chain to the observer placed in the center. The ``regular'' line is
for the FRW model. $\lambda$ is the affine parameter  and it grows with the
time which go from the left to the right. The vertical  lines mark the end and
the beginning of the holes.}
\label{zetacent}
\end{center}
\end{figure}

In this section we show the results for the observer in the center. As 
confirmed by Fig.\ \ref{zetacent}, the compensation in the redshift effect is
good: the photon is passing through an integer number of half holes.

\begin{figure}
\begin{center}
\includegraphics[width=17 cm]{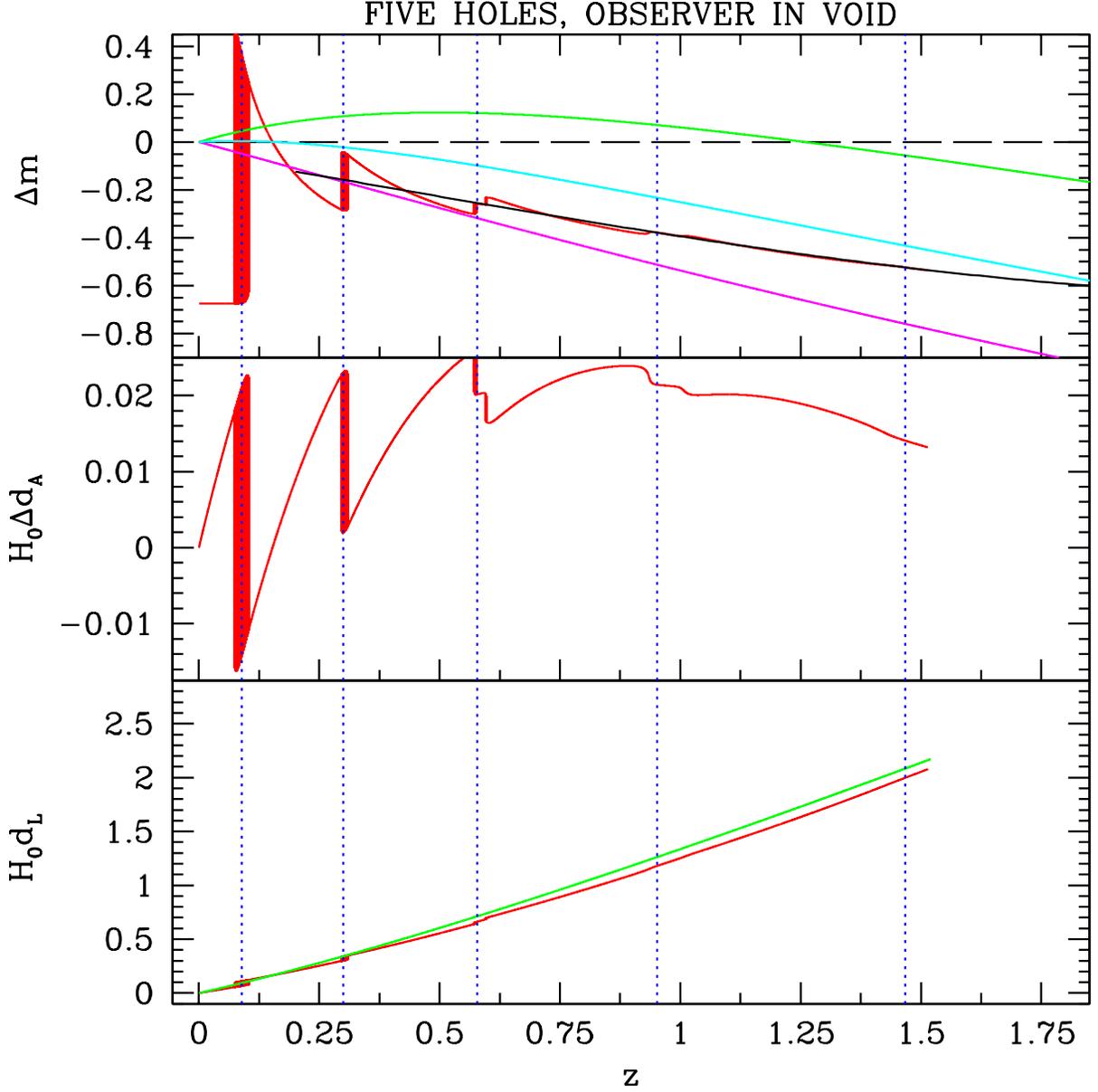}
\caption{The bottom panel shows the luminosity distance $d_L(z)$ in the
five-hole model (jagged curve) and the $\Lambda$CDM solution with
$\Omega_{M}=0.6$ and $\Omega_{DE}=0.4$ (regular curve). In the middle is the
change in the angular diameter distance, $\Delta d_A(z)$, compared to a
$\Lambda$CDM model with  $\Omega_{M}=0.6$ and $\Omega_{DE}=0.4$. On the top
panel the distance modulus in various cosmological models is shown.  The jagged
line is for the five-hole LTB model. The regular curves, from top to bottom are 
a $\Lambda$CDM model with  $\Omega_{M}=0.3$ and $\Omega_{DE}=0.7$, a
$\Lambda$CDM model with $\Omega_{M}=0.6$ and $\Omega_{DE}=0.4$, the best smooth
fit to the LTB model, and the EdS model.  The vertical lines mark the edges of
the five holes.}
\label{void}
\end{center}
\end{figure}

The results for the luminosity distance and the angular distance look worse as
shown in Fig.\ \ref{void}, but this is mainly due to the fact that now the
photon crosses half a hole less than in the previous cases and therefore it
undergoes less bending.

In this case the observer has no peculiar velocity compared to the FRW one:
this is a result of spherical symmetry.

\newpage
\ 
\newpage
\section{Conclusions} \label{conclusions}

The aim of this paper was to understand the role of large-scale non-linear 
cosmic inhomogeneities in the interpretation of observational data.  This
problem can be studied perturbatively, see for example Ref.\ \cite{kmr}.  Here,
instead, we focused on an exact (even if toy) solution, based on the 
Lema\^itre-Tolman-Bondi (LTB) model. This solution has been studied 
extensively in the literature  \cite{alnes0607, notari, alnes0602, celerier,
mansouri, flanagan, rasanen,  tomita, chung, nambu}. It has been shown that it can be
used to fit the observed $d_{L}(z)$  without the need of dark energy (for
example in \cite{alnes0602}).  To achieve this result, however, it is necessary
to place the observer at the center of a rather big underdensity. To overcome
this  fine-tuning problem we built a swiss-cheese model, placing the observer 
in the cheese and having the observer look through the swiss-cheese holes as
pictured in Fig.\ \ref{schizzo}.  A similar idea was at the basis of Refs.\
\cite{notari07, tetradis}.

Summarizing, we firstly defined the model in Section \ref{model}:  it is a
swiss-cheese model where the cheese is made of the usual FRW solution and the
holes are  made of a LTB solution. We defined carefully the free functions of
the LTB model in order to have a realistic (even if still toy) model and we
showed its dynamics in Section \ref{dynamics}.

Then, as anticipated in the Introduction, we focused on the effects of
inhomogeneities on photons. The observables on which we focused are the change
in redshift $\Delta z(\lambda)$ in angular-diameter distance $\Delta d_{A}(z)$,
in the luminosity distance-redshift relation $d_{L}(z)$, and in the distance
modulus $\Delta m(z)$.

We found that redshift effects are suppressed when the hole is small because of
a compensation effect acting on the scale of half a hole, due to spherical
symmetry: it is roughly due to the fact that $z'=H \propto \rho =
\rho_{\scriptscriptstyle FRW} + \delta \rho$ and we chose the density profile
in order to have $\langle \delta \rho \rangle=0$.  It is somewhat similar to
the screening among positive and negative charges.

However, we found interesting effects in the calculation of the angular
distance: the evolution of the inhomogeneities bends the photon path compared
to the FRW case. Therefore, inhomogeneities will be able (at least partly) to
mimic the effects of dark energy. We were mainly interested in making the
observer look through the swiss cheese from the cheese. However,  for a better
understanding, we examined also the case where the observer is inside the
hole.  We found bigger effects than those found in Refs.\ \cite{notari07,
tetradis}:  this could be due to the different model. Indeed,  Refs.\
\cite{notari07, tetradis} used smaller holes with a different 
initial-density/initial-velocity profile.

\clearpage

\acknowledgments{It is a pleasure to thank Alessio Notari and Marie-No\"elle C\'el\'erier
for useful discussions and suggestions.  V.M. acknowledges support
from ``Fondazione Ing. Aldo Gini'' and ``Fondazione Angelo Della Riccia.'' }




\appendix

\section{About the arbitrary functions in a LTB model} \label{arbifunc}

Here we illustrate, by means of an example, the choice of the arbitrary
functions in LTB models.  We are going to analyze the flat case.  Indeed we
have an analytical solution for it and this will help in understanding the
issues.

We said previously that there are three arbitrary functions in the LTB model: 
$\rho(r)$, $W(r)$ and $\bar{t}(r)$. They specify the position and velocities 
of the shells at a chosen time. In general, $\bar{t}$ depends on $r$; because
of the  absence of shell crossing it is possible to give the initial conditions
at different times for different shells labeled by $r$.

We start, therefore, by choosing the curvature $E(r)=(W^2(r)-1)/2$ to vanish,
which can be  thought as a choice of initial velocities $\dot{Y}$ at the time
$\bar{t}(r)$:
\begin{equation}
2 \, E(r)= \left. \dot{Y}^2-  
\frac{1}{3 \pi}\frac{M}{Y} \right|_{r, \, \bar{t}(r)} .
\end{equation}
For $E(r)=0$, the model becomes
\begin{equation}
ds^{2}=-dt^{2}+dY^{2}+Y^{2} d\Omega^{2} \;, 
\end{equation}
with solution
\begin{eqnarray} 
\label{flatso}
Y(r,t) & = & \left(\frac{3 \, M(r)}{4 \pi}\right)^{1/3}[t-\hat{t}(r)]^{2/3}
\nonumber \\
\bar{\rho}(r,t) & = & [t - \hat{t}(r)]^{-2} \;, 
\end{eqnarray}
where
\begin{eqnarray} 
\label{uffa}
\hat{t}(r) & \equiv & \bar{t}(r) - \bar{\rho}^{-1/2}(r, \bar{t}(r) )
\nonumber \\
\bar{\rho}(r, \bar{t}(r) ) & = & \left. \frac{3 \, M(r)}{4 \pi} 
\frac{1}{Y^{3}} \right|_{r, \, \bar{t}(r)} \;.
\end{eqnarray}
The next step is to choose the position of the shells, that is, to choose 
the density profile. As far as $\bar{t}(r)$ is concerned, only the 
combination $\hat{t}(r)$ matters. This, however, is not true for $M(r)$, 
which appears also by itself in Eq.\ (\ref{flatso}).

Looking at Eq.\ (\ref{uffa}) we see that to achieve an inhomogeneous profile 
we can either assign a homogeneous profile at an inhomogeneous initial time, or
an inhomogeneous density profile at a homogeneous initial time, or both.
Moreover, if we assign the function $M(r)$, then we can use our freedom to 
relabel $r$ in order to obtain all the possible $\bar{t}(r)$.  So we see that
one of the three arbitrary functions expresses the gauge freedom.

In this paper we fixed this freedom by choosing $\bar{t}(r)=\bar{t}$ and 
$Y(r, \bar{t})=r$ in order to have a better intuitive understanding 
of the initial conditions.

\section{Sewing the photon path} \label{sewing}

\begin{figure}
\begin{center}
\includegraphics[width=13 cm]{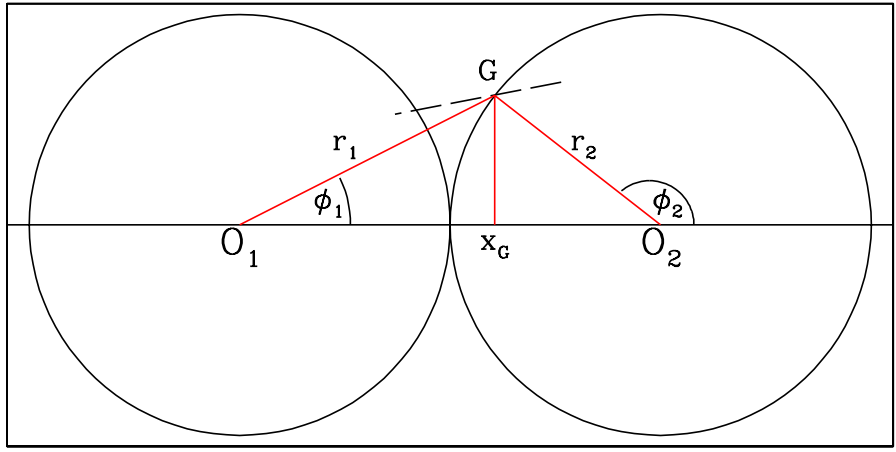}
\caption{Illustration of the procedure to calculate the transition between two
holes. The dashed line is a segment of the geodesic. $O_{1}$ and $O_{2}$
represent the two coordinate systems.}
\label{sew}
\end{center}
\end{figure}

In the Appendix we will demonstrate how to sew together the photon path between
two holes. We will always use center-of-symmetry coordinates, and therefore we
will move from the coordinates of $O_{1}$ to the ones of $O_{2}$ illustrated in
Fig.\ \ref{sew}. The geodesic near the contact point $G$ is represented by the
dashed line segment in Fig.\ \ref{sew}.

First, we want to find the value $\bar{\lambda}$ of the affine parameter for
which the photon is at $G$. This is found solving
\begin{eqnarray}
G_{2} & = & \left(r_{1}(\lambda) \cos \phi_{1}(\lambda)-2 r_{h},\; 
r_{1}(\lambda) \sin \phi_{1}(\lambda)\right) \nonumber \\
r_h^2 & = & x_{2 \, G}^{2}+y_{2 \, G}^{2}  .
\end{eqnarray}
These equations imply
\begin{equation}
r_{1}^{2}(\lambda)+3 r_{h}^{2}-4 r_{1}(\lambda)r_{h} \cos \phi_{1}(\lambda)=0 .
\end{equation}
Then we can give the initial conditions for the second hole:
\begin{eqnarray}
q_{2}(\bar{\lambda}) & = & q_{1}(\bar{\lambda}) \nonumber \\
t_{2}(\bar{\lambda}) & = & t_{1}(\bar{\lambda}) \nonumber \\
r_{2}(\bar{\lambda}) & = & r_{h} \nonumber \\
\phi_{2}(\bar{\lambda}) & = & \arccos(x_{2\, G}/r_{h}) .
\end{eqnarray}
Finally, we need the constant $c_{\phi}$, a sort of constant angular momentum
density. Repeating the procedure of Sect.\ \ref{ciccio} for the first hole we
find
\begin{equation}
c_{2\, \phi}= \sin \alpha_{2} \;  q_{1}(\bar{\lambda}) \; 
\left. Y_{2}\right|_{\bar{\lambda}} .
\end{equation}
Only $\alpha_{2}$ is missing. One way to find it is to calculate the inner
product in $O_{1}$ coordinates of the geodesic with the normalized spatial
vector parallel to $\overline{O_{2}G}$ (see Fig.\ \ref{sew}).

\end{document}